\documentclass[journal, twoside]{IEEEtran}
\usepackage{indentfirst}
\usepackage{graphicx}
\usepackage{amsmath}
\usepackage{amssymb}
\usepackage{amsfonts}
\usepackage{mathrsfs}
\usepackage{leftidx}
\usepackage{color}
\usepackage{amsmath}
\usepackage{arydshln}
\usepackage{amsthm}
\usepackage{ragged2e}
\usepackage{cite}
\usepackage{enumerate}
\usepackage{longtable}
\usepackage{float}
\usepackage{stfloats}
\usepackage{hyperref}
\usepackage{algpseudocode}
\usepackage{algorithm}
\usepackage[caption=false,font=footnotesize]{subfig}
\usepackage{tabularx}
\usepackage{makecell}
\usepackage{url}

\usepackage{multirow} 
\usepackage{booktabs}
\theoremstyle{plain}

\usepackage{caption}
\usepackage{multirow}   
\usepackage{array}      

\newcolumntype{P}[1]{>{\raggedright\arraybackslash\footnotesize}m{#1}}
\newcolumntype{A}[1]{>{\centering\arraybackslash\footnotesize}m{#1}}
\newcolumntype{M}[1]{>{\centering\arraybackslash}m{#1}}

\usepackage[table,usenames,dvipsnames]{xcolor}
\definecolor{aa}{RGB}{175,238,238}
\definecolor{bb}{RGB}{255,255,255}

\usepackage{bm}
\usepackage{makecell}

\begin{document}

\title{Semantic Radio Access Networks: 
Architecture, State-of-the-Art, and Future Directions}

\author{Rui Meng, \textit{Member, IEEE,} Zixuan Huang, Jingshu Yan, Mengying Sun, 
\textit{Member, IEEE,} Yiming Liu, \textit{Member, IEEE,} Chenyuan Feng, \textit{Member, IEEE,} Xiaodong Xu, \textit{Senior Member, IEEE,} Zhidi Zhang, Song Gao, 

Ping Zhang, \textit{Fellow, IEEE,} and Tony Q. S. Quek, \textit{Fellow, IEEE}

\thanks{
\textit{Rui Meng, Zixuan Huang, and Jingshu Yan contributed equally to this work and should be considered
co-first authors. Co-corresponding authors: Rui Meng, Mengying Sun, and Xiaodong Xu.}

Rui Meng, Jingshu Yan, Zixuan Huang, Mengying Sun, Yiming Liu, Xiaodong Xu, Zhidi Zhang, Song Gao, and Ping Zhang are with the State Key Laboratory of Networking and Switching Technology, Beijing University of Posts and Telecommunications, Beijing 100876, China (e-mail: buptmengrui@bupt.edu.cn; h2100zx@bupt.edu.cn; yaningshu0213@bupt.edu.cn; smy\_bupt@bupt.edu.cn; liuyiming@bupt.edu.cn; xuxiaodong@bupt.edu.cn; 2639134068@bupt.edu.cn; wkd251292@bupt.edu.cn; pzhang@bupt.edu.cn).

Chenyuan Feng is with Department of Computer Science, University of Exeter, EX4 4QF Exeter, U.K. (e-mail: c.feng@exeter.ac.uk).

Tony Q. S. Quek is with the Singapore University of Technology and Design, Singapore (e-mail: tonyquek@sutd.edu.sg).

}}

\maketitle

\begin{abstract}
Radio Access Network (RAN) is a bridge between user devices and the core network in mobile communication systems, responsible for the transmission and reception of wireless signals and air interface management.
In recent years, Semantic Communication (SemCom) has represented a transformative communication paradigm that prioritizes meaning-level transmission over conventional bit-level delivery, thus providing improved spectrum efficiency, anti-interference ability in complex environments, flexible resource allocation, and enhanced user experience for RAN. However, there is still a lack of comprehensive reviews on the integration of SemCom into RAN. Motivated by this, we systematically explore recent advancements in Semantic RAN (SemRAN).
We begin by introducing the fundamentals of RAN and SemCom, identifying the limitations of conventional RAN, and outlining the overall architecture of SemRAN. Subsequently, we review representative techniques of SemRAN across physical layer, data link layer, network layer, and security plane. Furthermore, we envision future services and applications enabled by SemRAN, alongside its current standardization progress. Finally, we conclude by identifying critical research challenges and outlining forward-looking directions to guide subsequent investigations in this burgeoning field.

\end{abstract}


\section{Introduction}

\subsection{Motivation}
As a core pillar of mobile communication systems, the Radio Access Network (RAN) assumes critical functions in connecting user equipment to the core network, encompassing capabilities such as user access management, wireless signal transmission, and optimized resource allocation \cite{chen2024evolution}. Currently, 5G-RAN has achieved integration of multiple technologies, including millimeter wave (mmWave), massive Multi-Input Multi-Output (MIMO), beamforming, network slicing, and Mobile Edge Computing (MEC), while supporting dual-mode operation in both standalone (SA) and non-standalone (NSA) architectures. This aims to satisfy the requirements of three typical scenarios: Ultra-Reliable Low-Latency Communications (URLLC), enhanced Mobile Broadband (eMBB), and massive Machine-Type Communications (mMTC) \cite{parvez2018survey}. However, in the evolution toward 6G and future wireless networks, RAN must achieve breakthrough improvements in core metrics such as intelligence level, spectral efficiency, and latency \cite{wang2023road}. Traditional evolutionary paths relying on increasing antenna counts, expanding spectrum resources, and enhancing power efficiency are no longer sufficient to meet development needs, necessitating breakthroughs in bottleneck areas \cite{polese2023understanding}.

In recent years, the deep integration of Artificial Intelligence (AI) and RAN has emerged as an innovative direction jointly explored by academia and industry \cite{khan2023ai}. By restructuring traditional RAN architectures, optimizing core techniques, and revolutionizing operational models, AI-RAN demonstrates significant advantages in enhancing network performance, energy efficiency, adaptability, and intelligence levels \cite{zhang2025comai}. Among these, by integrating Semantic Communication (SemCom) and RAN, Semantic RAN (SemRAN) stands out as one of the most representative technological directions. SemRAN utilizes the potential value of information in the semantic dimension, enabling a paradigm shift from ``transmitting syntactic bits'' to ``transmitting semantic models'' and establishing native semantic capabilities within RAN \cite{yang2022semantic,meng2025survey,lu2025important}. It has validated substantial advantages in areas such as interference mitigation, fading resistance, and service experience optimization. As a key enabling pathway toward \textit{intellicise (intelligent and concise)} wireless networks, SemRAN can provide critical support for the intelligent, efficient development of future mobile communication systems \cite{zhang2024intellicise,yining2024intellicise}.

To this end, this survey will focus on SemRAN, detailing its specific architecture. It will also comprehensively review key techniques, security and privacy protection methods, applications, and standardization efforts, aiming to inspire future research in this emerging field.

\subsection{Preliminaries and State-of-the-Art Works}

\begin{table*}
\caption{Comparison of Existing Survey/Tutorial papers on RAN and SemCom.}
\centering
\label{sotasurvey}

\rowcolors{2}{gray!15}{white}

\begin{tabular}{ccc M{1.6cm} M{1.8cm} M{1.7cm} M{1.6cm} M{2.0cm} c}
\hline
\addlinespace[0.5ex]
\textbf{Domain}  &  \textbf{Ref.} & \textbf{Year} & \textbf{Architecture} &
\textbf{Key Technologies} & \textbf{Security and Privacy}  &
\textbf{Applications}  & \textbf{Standardization}  & \textbf{Challenges} \\

\hline
\addlinespace[0.5ex]
RAN  & \cite{polese2023understanding} & 2023 & \textcolor{teal}{$\checkmark$} &
\textcolor{teal}{$\checkmark$} & \textcolor{teal}{$\checkmark$} &
\textcolor{red}{$\times$} & \textcolor{teal}{$\checkmark$} & \textcolor{teal}{$\checkmark$}\\

RAN  & \cite{brik2024explainable} & 2024 & \textcolor{teal}{$\checkmark$} &
\textcolor{teal}{$\checkmark$} & \textcolor{teal}{$\checkmark$} &
\textcolor{red}{$\times$} & \textcolor{teal}{$\checkmark$} & \textcolor{teal}{$\checkmark$}\\

RAN  & \cite{santos2025managing} & 2025 & \textcolor{teal}{$\checkmark$}  &
\textcolor{red}{$\times$}  & \textcolor{red}{$\times$}  &
\textcolor{red}{$\times$} & \textcolor{teal}{$\checkmark$}  &  \textcolor{teal}{$\checkmark$} \\

RAN  & \cite{alam2025comprehensive} & 2025 & \textcolor{teal}{$\checkmark$} &
\textcolor{teal}{$\checkmark$} & \textcolor{red}{$\times$} &
\textcolor{red}{$\times$} & \textcolor{red}{$\times$} & \textcolor{teal}{$\checkmark$} \\

RAN  & \cite{chen2024evolution} & 2025 & \textcolor{teal}{$\checkmark$} &
\textcolor{teal}{$\checkmark$} & \textcolor{red}{$\times$} &
\textcolor{teal}{$\checkmark$} & \textcolor{red}{$\times$} & \textcolor{teal}{$\checkmark$} \\

RAN  & \cite{herrera2025tutorial} & 2025 & \textcolor{teal}{$\checkmark$} &
\textcolor{red}{$\times$} & \textcolor{red}{$\times$} &
\textcolor{red}{$\times$} & \textcolor{teal}{$\checkmark$} & \textcolor{teal}{$\checkmark$} \\

SemCom  & \cite{yang2022semantic} & 2023 & \textcolor{teal}{$\checkmark$} &
\textcolor{teal}{$\checkmark$} & \textcolor{red}{$\times$} &
\textcolor{teal}{$\checkmark$} & \textcolor{red}{$\times$} &  \textcolor{teal}{$\checkmark$}\\

SemCom  & \cite{guo2024survey} & 2024 & \textcolor{teal}{$\checkmark$} &
\textcolor{red}{$\times$} & \textcolor{teal}{$\checkmark$} &
\textcolor{red}{$\times$} & \textcolor{red}{$\times$} &  \textcolor{teal}{$\checkmark$}\\

SemCom  & \cite{zhang2024intellicise} & 2024 & \textcolor{teal}{$\checkmark$} &
\textcolor{teal}{$\checkmark$} & \textcolor{red}{$\times$} &
\textcolor{teal}{$\checkmark$} & \textcolor{red}{$\times$} &  \textcolor{teal}{$\checkmark$}\\

SemCom  & \cite{won2024resource} & 2024 & \textcolor{teal}{$\checkmark$} &
\textcolor{teal}{$\checkmark$} & \textcolor{red}{$\times$} &
\textcolor{red}{$\times$} & \textcolor{red}{$\times$} &  \textcolor{teal}{$\checkmark$}\\

SemCom  & \cite{meng2025survey} & 2025 & \textcolor{teal}{$\checkmark$} &
\textcolor{teal}{$\checkmark$} & \textcolor{teal}{$\checkmark$} &
\textcolor{teal}{$\checkmark$} & \textcolor{red}{$\times$} &  \textcolor{teal}{$\checkmark$}\\

SemCom  & \cite{zhang2025resource} & 2025 & \textcolor{teal}{$\checkmark$} &
\textcolor{teal}{$\checkmark$} & \textcolor{red}{$\times$} &
\textcolor{red}{$\times$} & \textcolor{red}{$\times$} &  \textcolor{teal}{$\checkmark$}\\

SemRAN  & \cite{sun2025s} & 2025 & \textcolor{teal}{$\checkmark$} &
\textcolor{teal}{$\checkmark$} & \textcolor{red}{$\times$} &
\textcolor{red}{$\times$} & \textcolor{red}{$\times$} &  \textcolor{teal}{$\checkmark$}\\

SemRAN  & Ours & / & \textcolor{teal}{$\checkmark$} & \textcolor{teal}{$\checkmark$} &
\textcolor{teal}{$\checkmark$} & \textcolor{teal}{$\checkmark$} &
\textcolor{teal}{$\checkmark$} &  \textcolor{teal}{$\checkmark$}\\
\addlinespace[0.5ex]
\hline
\end{tabular}
\end{table*}

\subsubsection{Radio Access Networks (RANs)}
As a fundamental component of mobile communication systems, RAN is responsible for bridging user equipment with the core network and ensuring stable and efficient wireless access. Some representative survey papers on RANs are listed in Table \ref{sotasurvey}.
Polese \textit{et al.} \cite{polese2023understanding} introduce the architecture, interfaces, control mechanisms, and research progress of Open-RAN (O-RAN).
Brik \textit{et al.} \cite{brik2024explainable} systematically investigate the introduction mechanisms and deployment frameworks of explainable AI in AI-RAN. 
Chen \textit{et al.}  \cite{chen2024evolution} analyze the evolution of RAN from a high-level design perspective.
Alam \textit{et al.} \cite{alam2025comprehensive} mainly focus on the standardization status, architectural support, deployment schemes, and research challenges of O-RAN network slicing.

\subsubsection{Semantic Communications (SemCom)}
SemCom aims to enable efficient collaboration among communicating entities through the understanding of the semantics of information and knowledge, rather than merely performing data-level synchronization as in traditional communication systems \cite{yang2022semantic}. Table \ref{sotasurvey} also lists some representative survey papers on SemCom.
Yang \textit{et al.} \cite{yang2022semantic} unify the definitions and classification framework of SemCom and systematically review the key technologies and application prospects of SemCom in 6G.
Won \textit{et al.} \cite{won2024resource} conduct a survey on resource management, security, and privacy in SemCom, and develop a unified analytical framework encompassing semantic-aware resource allocation, security protection, and privacy preservation.
Meng \textit{et al.} \cite{meng2025survey} concentrate on the classification and analysis of security and privacy issues in SemCom.

\textit{Gaps and Motivations:}
Despite the fact that Sun \textit{et al.} \cite{sun2025s} propose a semantic-aware RAN architecture, extending SemCom research from predominantly single-link scenarios to multi-user, network-level systems, notably, a comprehensive understanding of the state-of-the-art in SemCom for RANs remains preliminary. 
To address this gap, we present a comprehensive survey that explores state-of-the-art SemRAN techniques for enhancing the performance of RAN.


\subsection{Key Contributions and Outline}

\subsubsection{Outline the Architecture of SemRAN}
We begin with a concise review of RAN and SemCom, followed by highlighting the advantages SemCom introduces to RAN. We then present the SemRAN architecture, which comprises four key components: the physical layer, data link layer, network layer, and security plane.

\subsubsection{Comprehensively Review Key Techniques}
Building on the proposed SemRAN architecture, we conduct a systematic review of representative techniques across these layers. The physical layer techniques includes semantic-based coding, semantic-aware Channel State Information (CSI) feedback, semantic-based beam management, and semantic-based multiple access (MA). The data link layer techniques includes semantic-based Hybrid Automatic Repeat reQuest (HARQ) and resource scheduling. The network layer techniques include semantic-based Knowledge Base (KB) and Age of Semantic Information (AoSI).

\subsubsection{Summarize Security Threats and Defense Technologies}

As an emerging paradigm, SemRAN faces novel security challenges. We categorize these threats into two main types: attacks targeting semantic models and attacks on semantic transmission. To address these, we survey defense strategies including data cleaning, robust enhancing, differential privacy, model compression, blockchain, cryptography technology, semantic steganography communication, covert semantic communication, and physical layer security.

\subsubsection{Prospect Applications and Provide the Standardization Progress}
We envision future services and applications enabled by SemRAN. Key services include Hyper Reliable and Low-Latency Communications (HRLLC), AI Generated Content (AIGC), and Integrated Sensing, Communication, Computation, and Intelligence (ISCCI). Application domains span immersive communications, agent-to-agent communications, Space-Air-Ground-Sea Integrated Network (SAGSIN), intelligent healthcare, smart factory, and intelligent transportation. We also outline current standardization efforts for SemRAN, including China Communications Standards Association (CCSA), IMT-2030, and 3rd Generation Partnership Project (3GPP).

\subsubsection{Explore the Future Directions}
We analyze the challenges of SemRAN and provide the future research directions, including the theoretical framework, management of semantic models, hybrid semantic-bit coexisting communication, and security and privation protection.

\begin{figure}
    \centering
    \includegraphics[width=1\linewidth]{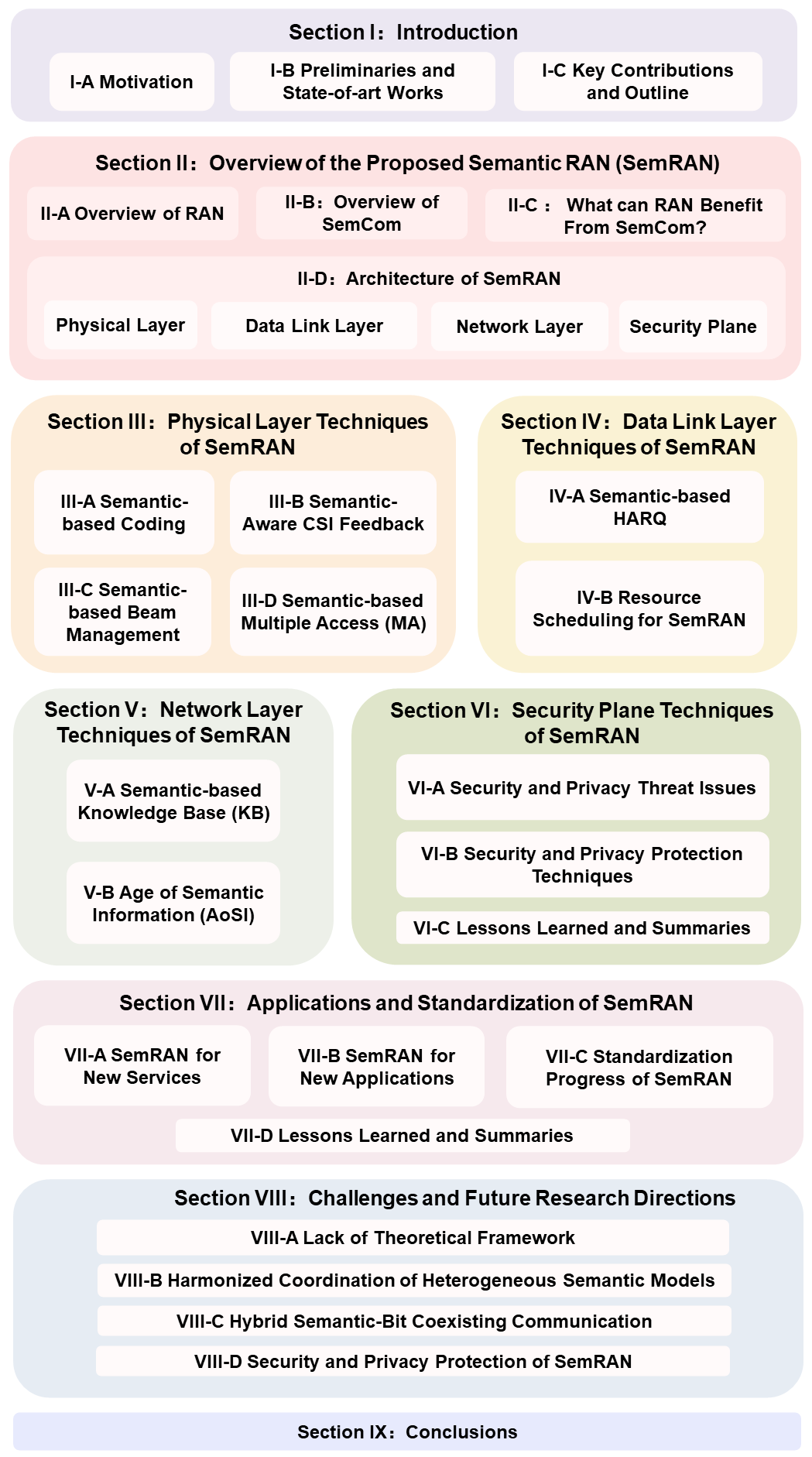}
    \caption{The structure of this paper.}
    \label{fig:Structure}
\end{figure}

\textit{Roadmap:}
The outline of this survey is illustrated in Figure \ref{fig:Structure}.
Section \ref{section2} introduces the architecture of SemRAN. Key technologies from the physical, data link, and network layers are then review in Sections \ref{section3}, \ref{section4}, and \ref{section5}. Security and privacy related aspects are discussed in Section \ref{section6}, while Section \ref{section7} goes into the applications and standardization of SemRAN. The challenges and potential directions are outlined in Section \ref{section8}. Finally, Section \ref{section9} concludes this survey.

\section{Overview of the Proposed Semantic RAN (SemRAN)}
\label{section2}

\subsection{Overview of RAN}

RAN provides the connection between user devices and the core network. Its main goal is to deliver stable and efficient wireless access by managing radio resources and supporting communication across different environments. A well-designed RAN should guarantee the coverage, capacity, and service reliability for diverse applications \cite{chen2024evolution}.
Traditional distributed RAN co-locates baseband processing and radio units, offering ease of deployment but limited flexibility. In contrast, Centralized RAN (C-RAN) separates remote radio units from centralized processing units, improving coordination and reducing processing cost \cite{habibi2019comprehensive}. More recently, O-RAN has emerged. It introduces open interfaces and disaggregated network functions. Through such design, equipment from different vendors can interoperate, and intelligent control becomes possible via a RAN intelligent controller \cite{polese2023understanding}. These architectures reflect the shift toward more scalable, open, and software-driven networks.
AI-RAN further employ virtualization, edge computing, and AI-driven optimization techniques to enhance network adaptability \cite{brik2024explainable,feng2025towards}.

\subsection{Overview of SemCom}


Compared with traditional syntactic communication, SemCom exhibits the following key distinctions:
\begin{itemize}
\item \textit{Processing Objects:} Syntactic communication treats bit streams as its processing unit, prioritizing the complete transmission of symbol-level data. In contrast, SemCom operates on semantic streams, emphasizing deep extraction and compression of information meanings \cite{cao2025importance,shi2025band}.
\item \textit{Processing Methods:} Syntactic communication adopts a separated source-channel design, whereas SemCom primarily employs Joint Source–Channel Coding (JSCC) \cite{bourtsoulatze2019deep} and generative coding techniques \cite{fan2025generative} to optimize semantic representation.
\item \textit{Bearing Forms:} Syntactic communication relies on fixed codebook mappings between codewords and symbols, while SemCom leverages semantic bases (Seb), models, and KBs to enable semantic feature modeling and evolution of semantic background knowledge \cite{fan2025kgrag}.
\item \textit{Evaluation Criteria:} Syntactic communication prioritizes bit-level error-free transmission metrics, e.g., bit error rate. SemCom, by contrast, focuses on task-oriented metrics such as semantic distortion, which directly measure the fidelity of meaning preservation \cite{nguyen2024distortion}.
\end{itemize}
Consequently, SemCom reduces communication overhead, enhances reliability, and supports intelligent information exchange, making it a promising candidate for integration into RANs \cite{sun2025s}.

\subsection{What can RAN Benefit From SemCom?}

\begin{itemize}
\item \textbf{Improvement of Spectrum Efficiency:}
SemCom only extracts and transmits the core semantics of information, reducing the occupation of spectrum by non critical data. Gao \textit{et al.} \cite{gao2025adaptive} verify that under the same transmission quality, compared with direct transmission of high-resolution video, SemCom can reduce the bandwidth consumption by 36.7\%.

\item \textbf{Anti-interference Ability in Complex Environments:}
SemCom realizes the joint adaptation of source information and channel information, which can achieve highly reliable transmission that adapts to dynamic changes of channel, and support low Signal-to-Noise Ratio (SNR) conditions \cite{teng2025conquering}. It has good anti-interference and anti-fading capabilities, further enhancing the seamless coverage expansion capability of RANs \cite{xu2023latent}.

\item \textbf{Flexible Resource Allocation:}
SemCom can allocate resources based on the importance and timeliness of semantic features \cite{zhang2025resource,xu2023task}. For example, in autonomous driving scenarios, the semantic priority of ``emergency brake command" is higher than that of ``car music", and higher bandwidth and lower latency wireless resources for it are allocated.

\item \textbf{Enhancement of User Experience:}
Semantic-enhanced perception, analysis, and evolution can autonomously and efficiently identify application tasks and understand user needs \cite{jiang2025large}. Therefore, it can dynamically adjust coding, resource allocation, and other strategies, ultimately enhancing the end-user experience \cite{wang2024feature}.

\end{itemize}

\subsection{Architecture of SemRAN}

\begin{figure}
    \centering
    \includegraphics[width=1\linewidth]{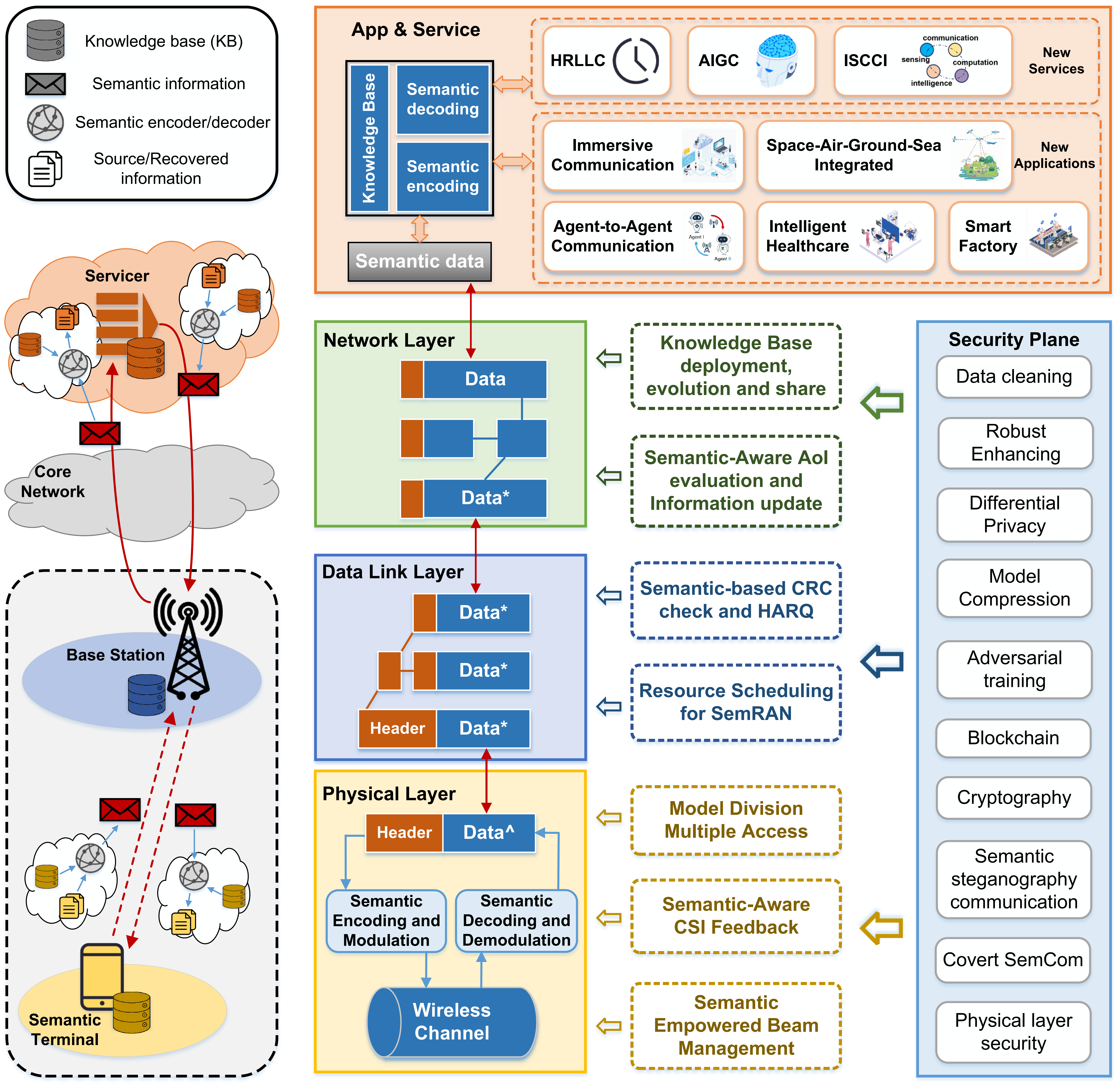}
    \caption{The proposed SemRAN architecture.}
    \label{fig:SemRAN}
\end{figure}

As illustrated in Figure \ref{fig:SemRAN}, SemRAN is composed of physical layer, data link layer, network layer, and security plane, which are described in detail as follows.
\subsubsection{Physical Layer}
The physical layer achieves the transition from ``bit transmission" to ``semantic transmission" through semantic-based signal processing and transmission optimization techniques, thereby improving spectral efficiency, reducing transmission delay, and enhancing adaptability to complex wireless environments. The representative physical layer techniques are as follows.
\begin{itemize}
\item \textbf{Semantic-based Coding:}
It extracts compact semantic features from raw signals and removes irrelevant bits, thus improving efficiency and robustness \cite{bourtsoulatze2019deep,xu2023latent}.
\item \textbf{Semantic-aware CSI Feedback:}
It combines the semantic features of CSI to lower feedback overhead and maintain stable feedback performance \cite{ren2025semcsinet,zheng2025semantic}.
\item \textbf{Semantic-based Beam Management:}
It uses environmental semantic features to guide beam decisions, reducing overhead and improving beam accuracy \cite{raha2025advancing,wu2024deep}.
\item \textbf{Semantic-based MA:} 
It separates users by exploiting the fact that different semantic models can not understand the semantic information generated by other models \cite{zhang2023model}. 
\end{itemize}

\subsubsection{Data Link Layer}
The data link layer achieves efficient utilization and fair allocation of resources through semantic-driven reliable transmission and resource management, improving the overall network throughput and user experience. The representative data link layer techniques are as follows.
\begin{itemize}
\item \textbf{Semantic-based HARQ:}
It prioritizes semantic feature preservation over exact bit recovery, thereby reducing overhead under dynamic radio conditions \cite{jiang2022deep,zheng2025semantic2}.
\item \textbf{Resource Scheduling:}
It allocates sensing, computing, communication, and knowledge resources to support semantic processing and task-oriented transmission \cite{zhang2025resource}.
\end{itemize}

\subsubsection{Network Layer}
The network layer achieves intelligent scheduling and optimization of network resources through semantic aware knowledge management, thus improving network capacity, reducing transmission latency, and supporting real-time communication requirements in complex business scenarios. The representative network layer techniques are as follows.
\begin{itemize}
\item \textbf{Semantic-based KB:}
It provides shared background knowledge to support semantic compression and reasoning-based correction \cite{fan2025kgrag}.
\item \textbf{AoSI:}
It measures the freshness of semantic information, thereby reducing unnecessary transmissions while preserving task-level effectiveness \cite{delfani2024semantics}.
\end{itemize}

\subsubsection{Security Plane}
SemRAN faces complex security threats \cite{guo2024survey}. Attackers target not only the semantic transmission process but also the semantic models \cite{meng2025survey}. To address these threats, SemRAN employs multi-dimensional and cross-layer defense strategies. Together, these methods ensure the confidentiality, integrity, authentication, access control, availability, privacy, and freshness secrecy.

\section{Physical Layer Techniques of SemRAN}
\label{section3}
\subsection{Semantic-based Coding}

Conventional separation-based coding schems suffers from the ``cliff effect," where slight channel degradation causes catastrophic decoding failures. To address this, SemRAN introduces semantic-based coding that integrates feature extraction, compression, and channel protection into an end-to-end learnable process. This paradigm significantly enhances transmission robustness by preserving essential semantics. We categorize these schemes into JSCC and generative coding, as illustrated in Table \ref{tab_sem_coding}.

\begin{table*}[t]
\centering
\renewcommand{\arraystretch}{1.1}
\caption{Representative Semantic-based Coding Schemes.}
\label{tab_sem_coding}

\renewcommand{\tabularxcolumn}[1]{m{#1}}

\begin{tabularx}{\textwidth}{|m{2cm}|m{3.5cm}|m{2.5cm}|X|}
\hline
\textbf{Sub-Category} 
& \textbf{Descriptions} 
& \textbf{Representative Schemes} 
& \textbf{Contributions} \\
\hline

\multirow{4}{*}{\textbf{JSCC}} 
& \multirow{4}{=}{Leverages deep neural networks to parameterize transmitter and receiver functions, mapping source data directly to continuous-valued channel symbols without explicit digital modulation.} 
& DeepJSCC \cite{bourtsoulatze2019deep} 
& Applies CNNs for end-to-end wireless image transmission, eliminating the separation between source and channel coding. \\
\cline{3-4}

& & NTSCC \cite{dai2022nonlinear} 
& Integrates a learnable entropy model for variable-length transmission and attention-guided masking for adaptive rate allocation. \\
\cline{3-4}

& & SwinJSCC \cite{zhang2023swinjscc} 
& Leverages Swin Transformers for multi-scale semantic modeling and spatial upsampling/refinement. \\
\cline{3-4}

& & JCM \cite{bo2024joint} 
& Tailors for digital SemCom using VAEs to map source data to discrete constellation points, enabling differentiability. \\
\hline

\multirow{3}{2.4cm}{\textbf{Generative Coding}} 
& \multirow{3}{=}{Employs advanced generative models to formulate semantic reconstruction as a conditional synthesis task using learned semantic priors.} 
& GAN-based \cite{lokumarambage2023wireless} 
& Deploys a conditional GAN architecture at the receiver to reconstruct realistic content from compact semantic segmentation maps transmitted over the channel. \\
\cline{3-4}

& & Diffusion-based \cite{xu2023latent} 
& Simulates physical channel interference via a forward Markov diffusion process and employs a U-Net-based reverse process to iteratively reconstruct original semantic features. \\
\cline{3-4}

& & LLM-based \cite{salehi2025llm} 
& Introduces the KG-LLM framework to compress textual data by extracting structured entity--relationship triples via knowledge graphs before encoding them into contextual vectors using LLMs. \\
\hline
\end{tabularx}
\end{table*}

\subsubsection{JSCC}
JSCC leverages deep neural networks to parameterize the transmitter and receiver functions, mapping high-dimensional source data directly into continuous-valued channel symbols without explicit digital modulation. Four typical JSCC schemes are as follows.
\begin{itemize}
\item 
\textbf{DeepJSCC:}
Bourtsoulatze \textit{et al.} \cite{bourtsoulatze2019deep} pioneer DeepJSCC, utilizing Convolutional Neural Networks (CNNs) as the encoder and decoder. This end-to-end mapping directly minimizes reconstruction distortion, demonstrating superior robustness over separation-based schemes in low-SNR regimes without explicit quantization.

\item 
\textbf{Nonlinear Transform Source-Channel Coding (NTSCC):}
To approach theoretical optimality, NTSCC integrates a learnable entropy model and attention-guided masking. It enables variable-length transmission and adaptive power allocation based on the semantic importance and structural complexity of image regions \cite{dai2022nonlinear}.

\item 
\textbf{SwinJSCC:}
SwinJSCC leverages a hierarchical architecture of swin transformer, with shifted window based self-attention to achieve efficient and multi-scale semantic modeling. It enables the encoder to compress images into rich, multi-scale representations, and allows the decoder to reconstruct them gradually through spatial upsampling and attention-based refinement \cite{zhang2023swinjscc}.

\item 
\textbf{Joint Coding-Modulation (JCM):}
JCM \cite{bo2024joint} learns a direct probabilistic mapping from source data to discrete constellation points, effectively bypassing the non-differentiability issues inherent in traditional digital modulation steps. 

\end{itemize}

\subsubsection{Generative Coding}
\begin{figure}
    \centering
    \includegraphics[width=1\linewidth]{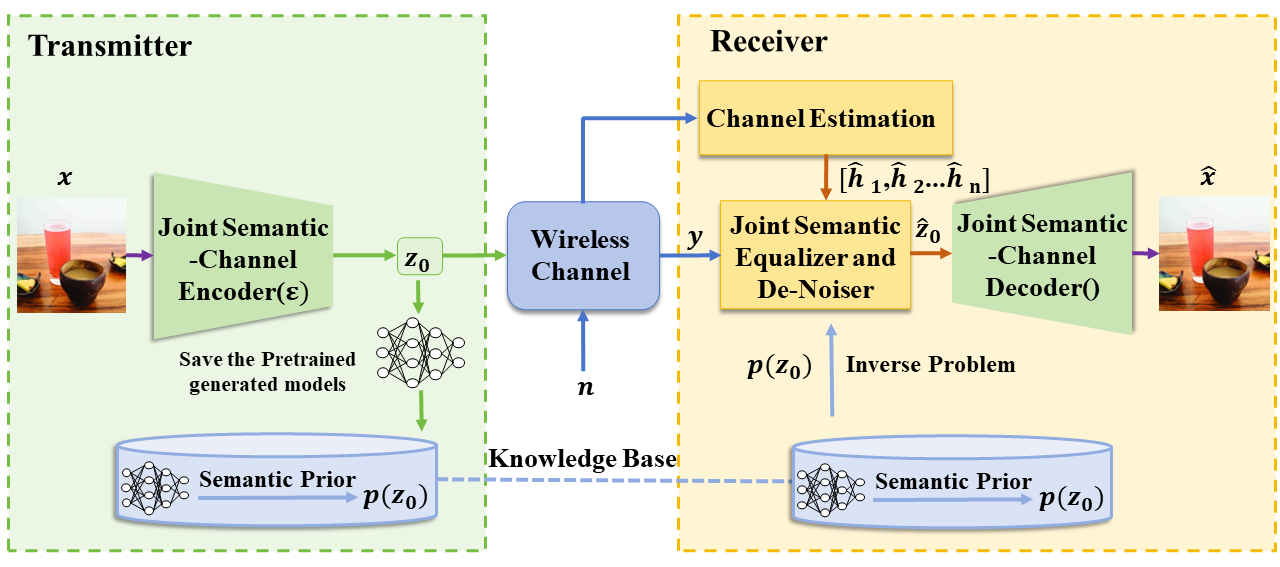}
    \caption{Illustration of the SP-EDNSC scheme \cite{xu2025semantic}, where semantic priors learned by latent diffusion models are exploited to perform channel-adaptive semantic equalizing and de-noising, enabling robust semantic reconstruction under fading channels.}
    \label{fig:Semantic-based Coding}
\end{figure}

Generative coding fundamentally redefines semantic decoding as a conditional generation task rather than deterministic reconstruction. By modeling the underlying distribution of source data, it leverages strong semantic priors to synthesize high-fidelity content from compact representations, prioritizing perceptual quality over pixel-wise alignment in bandwidth-limited scenarios.
\begin{itemize}
\item 
\textbf{Generative Adversarial Network (GAN)-based Coding:}
It incorporates an adversarial learning to treat the receiver as a conditional generator constrained by a discriminator to approximate the real data distribution. Lokumarambage \textit{et al.} \cite{lokumarambage2023wireless} propose a GAN-based semantic receiver for image transmission, where a generator synthesizes high-fidelity content from compressed semantic features while a discriminator provides adversarial feedback to optimize realism. 

\item 
\textbf{Diffusion-based Coding:}
It treats semantic decoding as an iterative denoising process, utilizing the reverse diffusion mechanism to progressively purify noisy signals into clean representations. Xu \textit{et al.} \cite{xu2023latent} propose diffusion-based coding to enable the SemRAN learning a robust probability distribution of channel noise, which facilitates adaptive signal cleaning during online inference without prior knowledge of the channel state. As illustrated in Figure \ref{fig:Semantic-based Coding}, Xu \textit{et al.} \cite{xu2025semantic} further propose a general SP-EDNSC scheme, which employs a latent diffusion model to remove channel effects from semantic features.

\item 
\textbf{Large Language Model (LLM)-based Coding:}
It is achieved by the advanced contextual reasoning capabilities of LLMs. Salehi \textit{et al.} \cite{salehi2025llm} employ an LLM encoder to compress the input into latent representations that capture essential semantic dependencies. The synchronized LLM at receiver reconstructs the semantic structure, which is subsequently refined by Bidirectional Encoder Representations from Transformers (BERTs) to correct errors and ensure high fidelity.
\end{itemize}

\subsection{Semantic-Aware CSI Feedback}

CSI feedback plays a crucial role in enabling channel-adaptive transmission in RAN. By providing the transmitter with channel knowledge, it allows dynamic adaptation of modulation, coding, and beamforming schemes, thereby enhancing both reliability and spectral efficiency. However, the explosive growth in the number of antennas have substantially increased the dimensionality of CSI \cite{wp5d2022future}, resulting in heavy feedback overhead and computational burden. It poses significant challenges to conventional CSI feedback schemes, including codebook-based \cite{qin2023review}, compressive sensing-based \cite{gao2018compressive}, and AI-based schemes \cite{guo2024deep}. 
Against this background, integrating SemCom into CSI acquisition and feedback processes offers a promising direction. Specifically, semantic-aware CSI feedback has the following advantages.
\begin{itemize}
\item \textbf{Significant Feedback Reduction:} 
It transmits only the essential task-related semantic representations instead of high-dimensional CSI data, substantially decreasing uplink overhead\cite{zhu2024semantic}.
\item \textbf{Improved Robustness and Generalization:}
It focuses on preserving features directly contributing to system-level tasks such as beam selection or link adaptation\cite{gao2023hybrid}.
\item \textbf{Task-Oriented Efficiency:}
Semantic representations, being abstract and compact, are more resilient to quantization errors, noise, and environmental changes, improving reliability across varying channel conditions\cite{zheng2025semantic}.
\end{itemize}

\begin{table*}[t]
\centering
\renewcommand{\arraystretch}{1.3} 
\caption{Representative Semantic-based CSI Feedback Methods.}
\label{tab_csifeedback}

\renewcommand{\tabularxcolumn}[1]{p{#1}}

\begin{tabularx}{\textwidth}{|p{2.8cm}|p{4.0cm}|p{2cm}|X|}
\hline
\textbf{Sub-Category} & \textbf{Descriptions} & \textbf{Representative Schemes} & \textbf{Contributions} \\
\hline
\multirow{2}{=}{\textbf{Side Information–based}}
& \multirow{2}{=}{Enhances CSI semantic representation by integrating auxiliary information that guides encoding or feature extraction.}
& LCFSC \cite{xie2024robust}
& Integrates CSI as side information into the semantic extraction process to enhance the robustness of semantic encoding
\\
\cline{3-4}
& & SemCSINet \cite{ren2025semcsinet}
& Injects CQI into the encoder to exploit semantic correlation between CSI and CQI, improving reconstruction accuracy.
\\
\hline
\multirow{2}{=}{\textbf{Adaptive Encoding–based}}
& \multirow{2}{=}{Optimizes CSI encoding based on factors such as semantic importance and predicted reconstruction quality.}
& Importance Aware \cite{zheng2025semantic}
& Allocates subcarrier resources adaptively and discards low-importance coefficients based on spatial characteristics.
\\
\cline{3-4}
& & SCAN \cite{zhang2024scan}
& Adaptively adjusts codeword length according to predicted reconstruction quality, enabling flexible semantic feedback.
\\
\hline
\multirow{2}{=}{\textbf{Semantic Label–based}}
& Represents CSI using semantic database labels shared by transmitter and receiver.
& \cite{zhu2024semantic}
& Constructs a label database via deep clustering and transmits only the label to reduce feedback overhead.
\\
\hline

\end{tabularx}
\end{table*}

Table \ref{tab_csifeedback} illustrates representative semantic-aware CSI feedback schemes, which can be divided into side information-based, adaptive encoding-based and semantic label-based methods.
\subsubsection{Side Information–based Methods}

These methods enhance CSI semantic representation by incorporating auxiliary information to guide the encoding process. Xie \textit{et al.} \cite{xie2024robust} propose the LCFSC framework, which integrates CSI as side information into the semantic extraction process and incorporates a non-invasive multi-head attention fusion module, thereby enhancing the robustness of semantic encoding. Ren \textit{et al.} \cite{ren2025semcsinet} design SemCSINet, injecting Channel Quality Indicator (CQI) into the CSI encoder to exploit the strong semantic correlation between CQI and CSI for improved reconstruction accuracy. Cao \textit{et al.} \cite{cao2023adaptive} propose a task-oriented semantic hiding method, embedding task-related semantic information into CSI via an information bottleneck formulation to balance feedback overhead and estimation performance.

\subsubsection{Adaptive Encoding–based Methods}

\begin{figure*}
    \centering
    \includegraphics[width=1\linewidth]{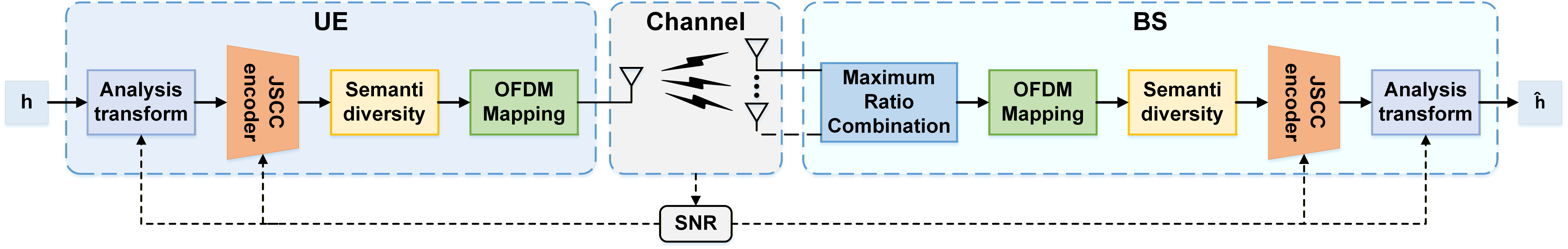}
    \caption{Illustration of the semantic-diversity–enabled CSI feedback scheme \cite{zheng2025semantic}, where semantic diversity is introduced into JSCC-based CSI feedback to selectively preserve critical semantic components and improve reconstruction reliability.}
    \label{fig:CSI Feedback}
\end{figure*}

These methods optimize the encoding and transmission of CSI based on factors such as semantic importance and predicted reconstruction quality. As illustrated in Figure \ref{fig:CSI Feedback}, Zheng \textit{et al.} \cite{zheng2025semantic} introduce an adaptive coding framework that allocates subcarrier resources and selectively discards low-importance symbols based on their spatial positions. Zhang \textit{et al.} \cite{zhang2024scan} present the SCAN model, which dynamically adjusts the CSI codeword length for each transmission instance depending on predicted reconstruction quality. Gao \textit{et al.} \cite{gao2023hybrid} introduce a hybrid knowledge–data–driven framework, jointly optimizing pilot design, CSI quantization, and semantic extraction, effectively implementing adaptive semantic coding for downstream tasks.

\subsubsection{Semantic Label–based Methods}

These methods leverage semantic labels from a pre-established database shared by transmitter and receiver \cite{zhu2024semantic}. The database is constructed via clustering and updated incrementally as new CSI is acquired. Only the label of the current CSI is transmitted, reducing feedback overhead while preserving task accuracy for semRAN.

\subsection{Semantic-based Beam Management}

Beam management builds and maintains proper transmit–receive beam pairs, reduces high-frequency path loss, and keeps the link stable for RANs \cite{giordani2018tutorial}. With beamforming, alignment, tracking, and prediction, RAN stays connected in dense and mobile environments.
However, large antenna arrays and narrow beams expand the search space and increase training needs, leading to high overhead and long delay \cite{imran2023environment}. Furthermore, fast channel changes, user mobility, and dynamic environments make the process even more difficult \cite{xue2024survey}. As a result, it is difficult for traditional exhaustive beam scanning to meet future requirements \cite{jeong2015random}. Many improved methods have been proposed, but they remain limited when antenna arrays become very large \cite{alkhateeb2014channel,jayaprakasam2017robust,heath2016overview,zhang2021learning}.
Against this background, integrating SemCom into beam management offers a promising direction. Specifically, semantic-based beam management provides the following advantages:

\begin{itemize}
\item \textbf{Lower Beam Management Overhead and Latency:} 
Semantic features keep only the information needed for beam decisions. This reduces the search space and avoids unnecessary processing. As a result, the overall beam management overhead and latency become much lower \cite{yang2023environment}.
\item \textbf{Better Adaptation to Dynamic Environments:}
Semantic information helps the system understand key environmental changes, such as mobility or blockage. With these cues, the link remains stable and the service quality improves, even in fast-varying scenarios \cite{raha2025advancing}.
\item \textbf{Higher Beam Accuracy:}
Because semantic features are closely related to the beamforming task, they guide the system toward more accurate beam choices. This improves alignment precision and enhances the overall beam performance \cite{raha2025advancing}.
\end{itemize}

\begin{table*}[t]
\centering
\renewcommand{\arraystretch}{1.3}
\caption{Representative Semantic-based Beam Management Methods.}
\label{tab_beam_management}

\renewcommand{\tabularxcolumn}[1]{p{#1}}

\begin{tabularx}{\textwidth}{|p{2cm}|p{4.0cm}|p{2.4cm}|X|}
\hline
\textbf{Sub-Category} & \textbf{Descriptions} & \textbf{Representative Schemes} & \textbf{Contributions} \\
\hline
\multirow{2}{=}{\textbf{Semantic-based Beamforming}}
&\multirow{2}{=}{Embeds high-level environmental or channel semantics into the beamforming optimization process to enhance robustness, efficiency, and task awareness. }
& SERBF \cite{raha2025advancing}
& Integrates task-relevant environmental semantics and semantic localization to enhance beamforming reliability.
\\
\cline{3-4}
& & JSCBF \cite{wu2024deep}
& Uses fused semantic features with hybrid networks to improve beamforming performance.
\\
\cline{3-4}
& & Energy-efficient Beamforming \cite{sun2025towards}
& Uses semantic descriptors to guide passive beam control and improve energy efficiency.
\\
\hline
\multirow{2}{=}{\textbf{Semantic-based Beam Prediction}}
& \multirow{2}{=}{Infers the optimal beam index from environmental or sensor semantics, reducing training overhead and latency.}
& COSCs Aided Prediction \cite{yang2023environment}
& Uses environmental semantics for joint beam and blockage prediction without pilot training.
\\
\cline{3-4}
& & SemQNet \cite{khan2025semqnet}
& Applies quantized semantic features for lightweight and low-cost beam index inference.
\\
\hline
\multirow{2}{=}{\textbf{Semantic-based Beam Selection}}
& Selects the optimal beam index from environmental semantics without exhaustive probing.
& Vision-based Selection \cite{wen2023vision}
& Uses semantic heatmaps from user images for accurate and low-overhead beam selection.
\\
\hline

\end{tabularx}
\end{table*}

Table \ref{tab_beam_management} summarizes representative semantic-based beam management schemes, which can be divided into semantic-based beamforming, beam prediction, and beam selection.

\subsubsection{Semantic-based Beamforming}

Semantic-based beamforming embeds high-level environmental or channel semantics into the beamforming optimization process to enhance robustness, efficiency, and task awareness. Raha \textit{et al.} \cite{raha2025advancing} design a semantic-empowered framework that incorporates task-relevant environmental semantics and semantic localization, achieving reliable operation in dynamic millimeter-wave scenarios. The framework further combines visual and positional sensing to enhance beamforming under mobility.Semantic information can also replace high-dimensional CSI. Wu \textit{et al.} \cite{wu2024deep} propose the JSCBF architecture, where fused semantic features guide semantic beamforming. The design uses both data-driven and model-driven networks and fuses their outputs to improve performance. Semantic priors also support energy-efficient beamforming. Metasurface-assisted holographic MIMO systems use semantic descriptors to guide passive beam control. Sun \textit{et al.} \cite{sun2025towards} show that stacked metasurfaces can improve semantic energy efficiency by reducing power consumption.

\subsubsection{Semantic-based Beam Prediction}

\begin{figure}
    \centering
    \includegraphics[width=1\linewidth]{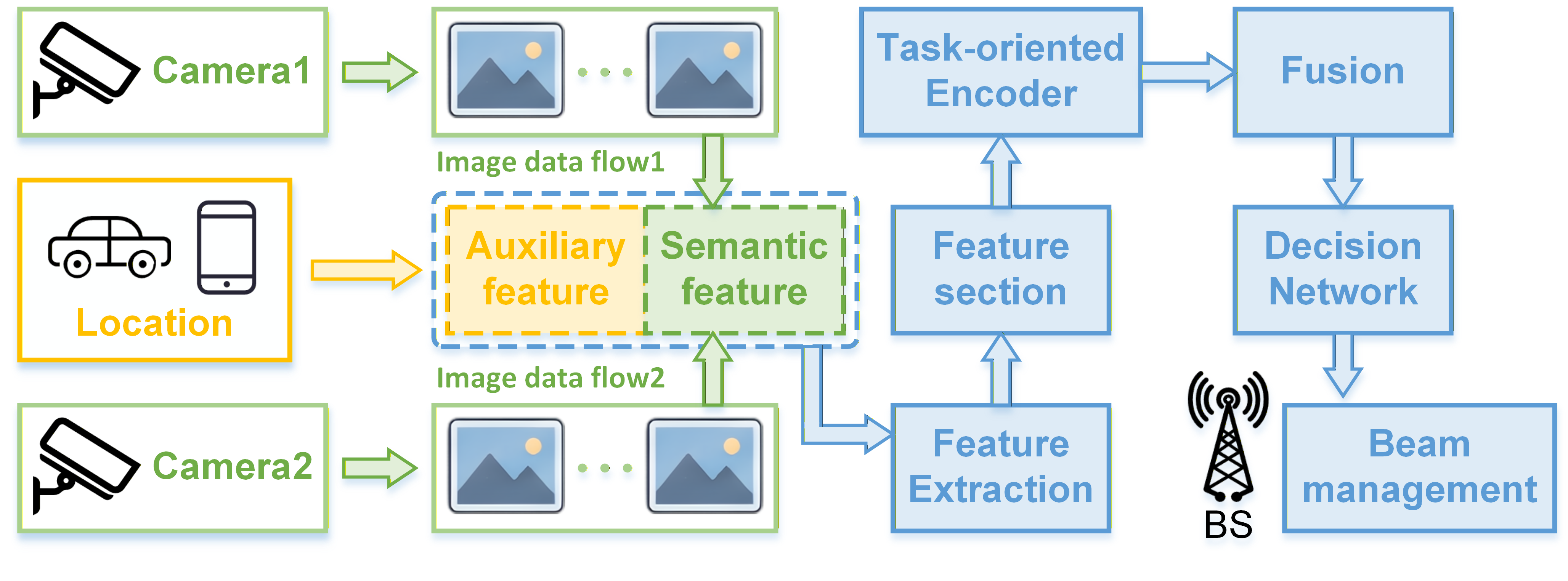}
    \caption{Illustration of the environment semantics–aided beam management framework \cite{yang2023environment}, where environment semantics from roadside cameras and auxiliary user features are selectively encoded and jointly exploited at the BS to support beam management.}
    \label{fig:Beam Management}
\end{figure}

Semantic-based beam prediction infers the optimal beam or beam index from environmental or sensor semantics, reducing training overhead and latency. Sun \textit{et al.} \cite{sun2023define} show that scatterer-based semantic descriptions enable accurate millimeter-wave beam prediction and reduce search cost. As illustrated in Figure \ref{fig:Beam Management}, Yang \textit{et al.} \cite{yang2023environment} extract key environmental semantics from sensory data and use them for joint beam and blockage prediction. This design avoids pilot training and beam scanning, achieving low latency and high reliability in dynamic conditions.Under energy constraints, semantic quantization is effective. Khan \textit{et al.} \cite{khan2025semqnet} propose SemQNet, which uses quantized semantic representations to reduce model size and infer beam indices with very low computational and communication cost, making it suitable for low-power platforms.

\subsubsection{Semantic-based Beam Selection}

This method selects the optimal beam index from environmental semantics without exhaustive probing. Wen \textit{et al.} \cite{wen2023vision} propose a vision-based framework that extracts semantic features from user images, including dominant scatterers and geometric layouts, and converts them into semantic heatmaps. A neural network uses these heatmaps for beam selection, improving accuracy while reducing traditional selection overhead.

\subsection{Semantic-based Multiple Access (MA)}

In RANs, common MA technologies include time division MA (TDMA), frequency division MA (FDMA), code division MA (CDMA), orthogonal frequency division MA (OFDMA), and non-orthogonal MA (NOMA). 
They divide resources in the time, frequency, code or space domain, so that multiple users can access the network at the same time without interfering with each other, thereby improving system capacity and connection density. 
However, with the increase in the number of users and the diversification of business needs, traditional MA faces the following challenges.

\begin{itemize}
\item \textbf{Rigidity of Resource Granularity:} 
The division of MA resources is fixed and it is difficult to adapt flexibly to the needs of heterogeneous services \cite{mao2021multiple}.
\item \textbf{Interference Management Complex:}
Especially in non-orthogonal scenarios, inter-user interference is significant and decoding complexity is high \cite{mao2021multiple}.
\item \textbf{Semantic Redundancy Not Exploited:}
Traditional MA focuses on bit-level transmission efficiency, ignoring the redundancy and relevance of the semantic layer \cite{zhang2023model}.
\end{itemize}

Under SemRAN, semantic-based MA shows unique advantages. By performing user differentiation and resource allocation in the modulation domain, semantic-based MA enables the semantic information to complete the initial separation and mapping at the physical layer, thereby improving the semantic transmission efficiency and robustness of SemRAN \cite{frauendorf2023the}. Besides, by leveraging different semantic models, semantic-based MA can map user source information to different model information spaces, thereby providing a knowledge base for distinguishing multi-user semantic information and supporting transmission from more devices \cite{wu2025joint}. 
Table \ref{tab_sbma} illustrates representative semantic-based MA schemes, which can be divided into independent semantic mapping-based, joint training-based, and orthogonality-constrained methods.

\begin{table*}[t]
\centering
\renewcommand{\arraystretch}{1.25}
\caption{Representative Semantic-based MA Schemes.}
\label{tab_sbma}
\begin{tabular}{|p{2.5cm} |p{5cm}|p{2.8cm}| p{4.4cm}|}
\hline
\textbf{Sub-Category} & \textbf{Descriptions} & \textbf{Representative Schemes}      & \textbf{Contributions} \\
\hline
\multirow{2}{2.5cm}{\textbf{Independent Mapping}} &  \multirow{2}{*}{\begin{tabular}[c]{@{}l@{}}Assigns each user an independently\\  trained semantic model, assumes \\ natural separability\end{tabular}}  & MDMA \cite{zhang2023model} & Allocates a unique pair of semantic encoder-decoder pairs between each user \\ \cline{3-4}
&     & DeepMA \cite{zhang2024deepma}       & Introduces a semantic symbol vector that is orthogonal to each other between users  \\
\hline
\multirow{2}{*}{\textbf{Joint Training}}      & \multirow{2}{*}{\begin{tabular}[c]{@{}l@{}}Trains multiple semantic models \\ jointly under interference to \\ enforce mutual exclusivity\end{tabular}} & JDASD \cite{wu2025joint}            & Proposes a joint training framework based on GAN \\ \cline{3-4}
 &  & Fusion-MDMA \cite{wu2023fusion}     & Designs a semantic fusion module to merge the functions of multiple users. \\
 \hline
\textbf{Orthogonality-Constrained}            & Defines and enforces semantic orthogonality via model structure or parameter diversity                    & O-MDMA \cite{liang2024orthogonal}   & Defines the semantic orthogonal signal as a signal generated by different semantic models \\
\hline
\end{tabular}
\end{table*}

\subsubsection{Independent Mapping-based Methods}
\begin{figure}
    \centering
    \includegraphics[width=1\linewidth]{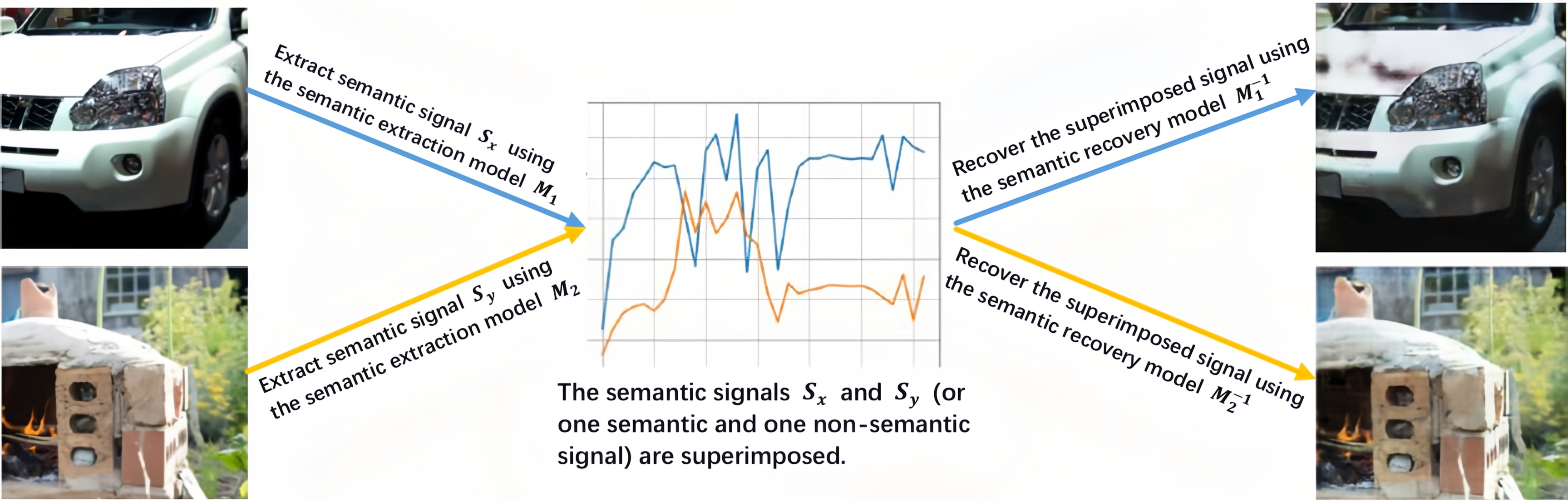}
    \caption{Illustration of mode-division multiple access \cite{zhang2023model}, where semantic signals extracted by different semantic models are superimposed and then recovered using the corresponding semantic recovery models.}
    \label{fig:MDMA}
\end{figure}

These schemes assign pre-trained semantic models to different users without considering inter-model interference during training. Semantic separation is achieved by leveraging inherent differences in model structure or parameters. As illustrated in Figure \ref{fig:MDMA}, Mode division multiple access (MDMA) \cite{zhang2023model} allocates a unique pair of semantic encoder-decoder pairs between each user. It assumes that the semantic signals of different models are sufficiently different to allow the receiver to separate. Besides, DeepMA \cite{zhang2024deepma} introduces a semantic symbol vector that is orthogonal to each other between users, and uses a decoder to reconstruct each user's data from the superimposed semantic symbol vector.

\subsubsection{Joint Training-based Methods}
These methods simultaneously train multiple semantic models under multi-user interference to explicitly optimize mutual exclusion and robustness. Wu \textit{et al.} \cite{wu2025joint} propose a joint training based on GAN, where ach semantic model acts as a generator and guides the model to encode information into a mutually exclusive semantic space through private discriminators and global discriminators. Wu \textit{et al.} \cite{wu2023fusion} propose Fusion-MDMA, where a semantic fusion module is employed to merge the functions of multiple users. The neural network adaptively weights semantic features according to channel conditions to improve the recovery accuracy in image transmission tasks. 

\subsubsection{Orthogonality-Constrained Methods}
These methods clearly define and enforce semantic orthogonality between models, usually using statistical indicators or structural constraints. For example, O-MDMA \cite{liang2024orthogonal} defines the semantic orthogonal signal as a signal generated by different semantic models and unable to decode each other's output. The similarity of quantitative models is used to guide model selection or training to ensure orthogonality.

\section{Data Link Layer Techniques of SemRAN}
\label{section4}
\subsection{Semantic-based HARQ}
HARQ combines forward error correction with automatic retransmission, which is important for ensuring air-interface transmission reliability in RAN. 
In SemRAN, the emphasis shifts from bit-level reconstruction to the preservation of semantic features, contextual dependencies, and task-level usability. While conventional HARQ ensures bit-level reliability through CRC-based error detection and fixed or limited variable-rate retransmission, it lacks the ability to assess semantic importance or leverage residual informative content from corrupted packets, which leads to unnecessary overhead under low-SNR conditions\cite{jiang2022deep,zhou2022adaptive,zheng2025semantic2,sheng2024semantic}.
Table~\ref{tab:semantic_transmission_schemes} lists the key semantic-based HARQ schemes proposed in recent literature, categorized by transmission modality.

\begin{table*}[t]
\centering
\caption{Representative Semantic-based HARQ Schemes}
\label{tab:semantic_transmission_schemes}
\renewcommand{\arraystretch}{1.4}

\begin{tabular}{
|>{\centering\arraybackslash}m{2.0cm}
|>{\RaggedRight\arraybackslash}m{4.4cm}
|>{\centering\arraybackslash}m{2.6cm}
|>{\RaggedRight\arraybackslash}m{5.8cm}|
}
\hline
\textbf{Sub-category} &
\textbf{Description} &
\textbf{Representative Schemes} &
\textbf{Contributions} \\
\hline

\multirow{2}{2.0cm}{\centering\makecell{Text Modality}} &
\multirow{2}{4.5cm}{\RaggedRight
Transmits natural language sentences; focuses on sentence meaning rather than bit-exact recovery.} &
SCHARQ \cite{jiang2022deep} &
Replaces CRC with Sim32 similarity detection, reducing sentence error rate and average transmitted bits under high BER. \\
\cline{3-4}

& &
IK-HARQ \cite{zhou2022adaptive} &
Proposes IK-HARQ with policy-based dynamic coding-rate selection and a single-encoder architecture to reduce retransmission overhead. \\
\hline

Image Modality &
Transmits images requiring fine-grained consistency of texture and color using explicit shared semantic bases. &
SAFG-HARQ \cite{zheng2025semantic2} &
Uses explicit semantic bases with fine-grained selective retransmission to preserve visual consistency and enhance reliability. \\
\hline

Multi-modality &
Transmits intermediate feature tensors across multiple modalities, where task performance depends on preserving critical semantic regions. &
SimHARQ-I/II \cite{sheng2024semantic} &
Employs importance-aware feature protection and task-driven semantic CRC to enhance perception accuracy in low-SNR environments. \\
\hline
\end{tabular}

\vspace{0.2cm}
\end{table*}

\subsubsection{Semantic-based HARQ for Text Modality}
\begin{figure}
    \centering
    \includegraphics[width=1\linewidth]{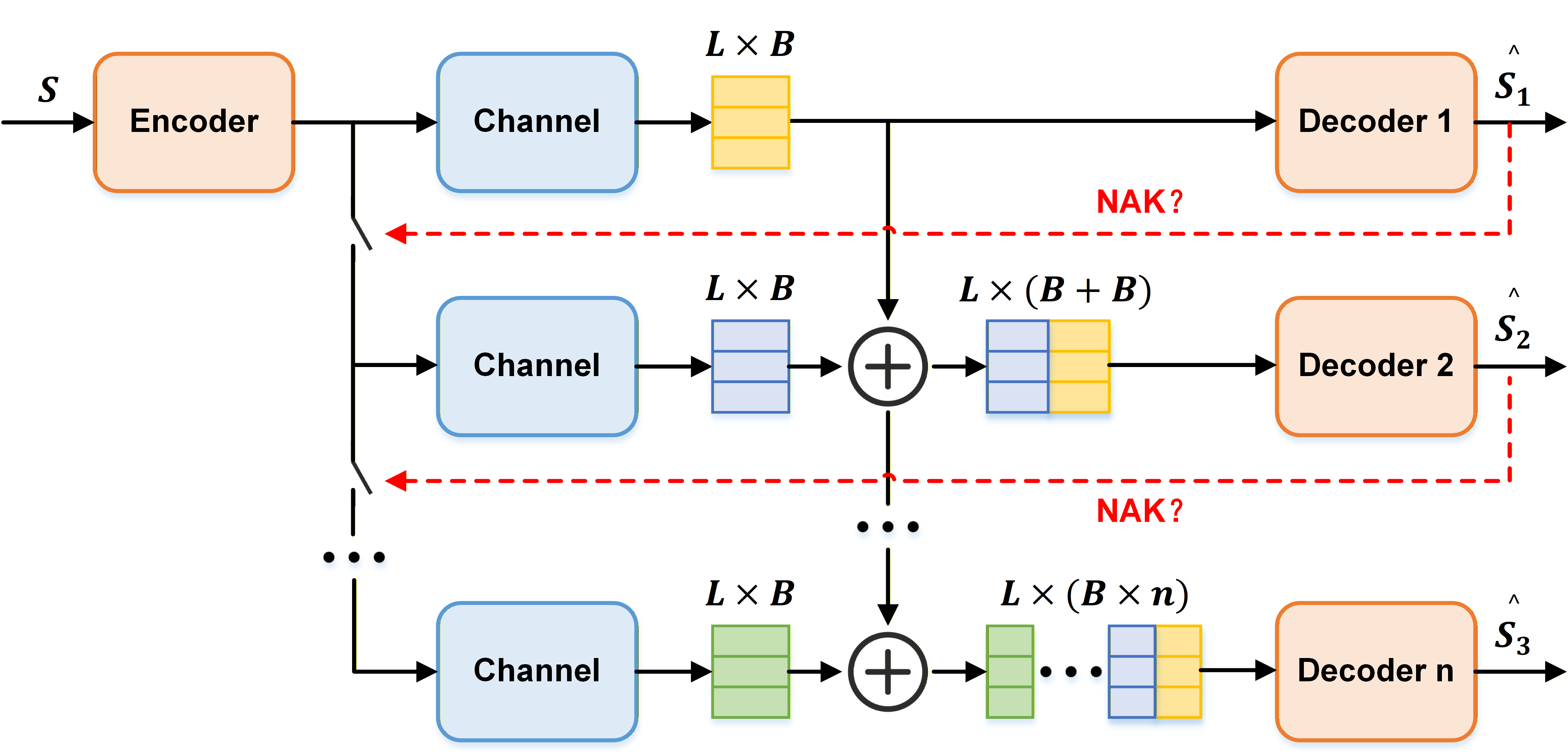}
    \caption{Illustration of the semantic IK-HARQ with multi-decoders \cite{zhou2022adaptive}, where a shared semantic encoder and multiple decoders enable progressive semantic refinement through incremental knowledge across retransmissions.}
    \label{fig:Semantic-based HARQ}
\end{figure}

Jiang \textit{et al.}~\cite{jiang2022deep} propose the end-to-end SCHARQ framework that combines variable-length JSCC with HARQ. They replace conventional CRC with a Transformer-based similarity detection network named Sim32. Simulation results show reduced sentence error rate and fewer average transmitted bits under high bit-error-rate conditions.
As illustrated in Figure \ref{fig:Semantic-based HARQ}, Zhou \textit{et al.}~\cite{zhou2022adaptive} propose incremental knowledge-based HARQ (IK-HARQ) that reuses residual semantic information from previous transmissions. A policy network selects the coding rate based on sentence semantic complexity and channel state.
    
\subsubsection{Semantic-based HARQ for Image Modality}
Zheng \textit{et al.}~\cite{zheng2025semantic2} propose an explicit Seb shared between transmitter and receiver. They design Semantic-based fine-grained HARQ (SAFG-HARQ) that performs error localization at the Seb level using contextual correlations and hypothesis-testing-based deviation compensation. 
The simulation results show that the transmission overhead is reduced by over 60\%, and LPIPS is improved by over 20\%.

\subsubsection{Semantic-based HARQ for Multi-modality}
Sheng \textit{et al.}~\cite{sheng2024semantic} propose an importance map to identify critical elements in feature tensors. Two HARQ variants are presented: SimHARQ-I using chase combining and SimHARQ-II using incremental redundancy. Simulations over time-varying multipath fading channels show higher perception accuracy and throughput compared with conventional HARQ.

\subsection{Resource Scheduling for SemRAN}

Resource scheduling is responsible for allocating limited resources, such as bandwidth, power, and computational capacity, to optimize RAN performance metrics like throughput, latency, and energy efficiency. In SemRAN, resource scheduling faces unique challenges due to the shift from bit-level transmission to semantics-level processing. It introduces additional complexities, including the need to handle semantic extraction, compression, and recovery while maintaining task-oriented quality of service \cite{zhang2025resource}. 
As illustrated in Table \ref{tab:resource_classification}, the optimized resource types include sensing, computing, communication, and knowledge resources.

\begin{itemize}
\item \textbf{Sensing Resource:} It is responsible for acquiring environmental and contextual data, such as radar waveforms and visual inputs, to enable semantic extraction \cite{lu2024semantic}.

\item \textbf{Computing Resource:} It handles neural network inference and the execution of semantic coders. Representative works focus on optimizing computation offloading and reducing latency in satellite and terrestrial edge environments \cite{zheng2024semantic,cang2023online,zhao2023joint}.

\item \textbf{Communication Resource:} It allocates bandwidth, sub-channels, and power for transmitting compressed semantic symbols. Key studies address semantic-aware power management \cite{xu2025generative}, bandwidth allocation and user association \cite{wang2024adaptive}, and joint bandwidth optimization \cite{xia2023joint}.

\item \textbf{Knowledge Resource:} It exploits shared background knowledge or pre-trained foundation models to improve compression efficiency. Representative studies include generative foundation models \cite{xu2025generative} and task-oriented adaptive compression \cite{liu2023adaptable}.
\end{itemize}

\begin{table}[t]
\centering
\caption{Classification of Resource Scheduling in SemRAN}
\label{tab:resource_classification}
\begin{tabular}{cccccc}
\toprule
\textbf{Type} & \textbf{Ref.} & \textbf{Sens.} & \textbf{Comp.} & \textbf{Comm.} & \textbf{Know.} \\
\midrule
\multirow{2}{*}{Single}  
  & \cite{xu2025generative}   & -- & -- & \checkmark & -- \\ 
  & \cite{lu2024semantic}     & \checkmark & -- & -- & -- \\
\midrule
\multirow{7}{*}{Dual}  
  & \cite{ding2023joint}      & -- & \checkmark & \checkmark & -- \\
  & \cite{zheng2024semantic}  & -- & \checkmark & -- & \checkmark \\
  & \cite{xia2023joint}       & -- & -- & \checkmark & \checkmark \\
  & \cite{zhao2023joint}      & -- & \checkmark & \checkmark & -- \\
  & \cite{liu2023adaptable}   & -- & -- & \checkmark & \checkmark \\ 
  & \cite{wang2024adaptive}   & -- & -- & \checkmark & \checkmark \\
  & \cite{hu2023semantic}                & -- & -- & \checkmark & \checkmark \\  
\midrule
\multirow{4}{*}{Multi}  
  & \cite{cang2023online}     & -- & \checkmark & \checkmark & \checkmark \\
  & \cite{minani2024qosem}       & -- & \checkmark & \checkmark & \checkmark \\  
  & \cite{hu2025resource}                & \checkmark & -- & \checkmark & \checkmark \\  
  & \cite{adhikary2024holographic}                & \checkmark & \checkmark & \checkmark & -- \\  
\bottomrule
\end{tabular}
\vspace{0.1cm}
\\ \small{
 \checkmark : Explicitly optimized variable. \\
-- : Not jointly optimized.
}
\end{table}








\section{Network Layer Techniques of SemRAN}
\label{section5}
\subsection{Semantic-based Knowledge Base (KB)}
Unlike static databases, the semantic KB stores semantic logic and relationships, facilitating a shift from data-driven to reasoning-driven operations \cite{LessDataMoreKnowledge}.
Table \ref{tab:kb_schemes} summarizes representative schemes, comparing their forms, mechanisms, and contributions.

\begin{table*}[htbp]
    \centering
    \caption{Summary of Semantic-based Knowledge Base Schemes}
    \label{tab:kb_schemes}
    
    \renewcommand{\arraystretch}{1.5}
    
    \begin{tabular*}{\textwidth}{
    |>{\centering\arraybackslash}m{2.2cm}
    |>{\raggedright\arraybackslash}m{3.2cm}
    |>{\centering\arraybackslash}M{1.95cm}
    |>{\raggedright\arraybackslash}m{9cm}|
    }
    \hline
    \textbf{Sub-Category} & 
    \textbf{Descriptions} & 
    \textbf{Representative Schemes} & 
    \textbf{Contributions} \\
    \hline

    \multirow{4}{2.2cm}{\centering Architecture and Representation} 
    & Generative Architecture 
    & \cite{KnowledgeBaseEnabledGenerative} 
    & Propose architecture with Source, Task, and Channel sub-bases to address cross-modal and cross-task challenges. \\
    \cline{2-4}

    & Hierarchical Framework 
    & \cite{UnifiedHierarchicalSKB} 
    & Employ horizontal and vertical construction to maximize representation space and explore task correlations. \\
    \cline{2-4}

    & Explicit Semantic Bases 
    & \cite{ExplicitSemanticBase} 
    & Utilize clustering-generated explicit semantic bases to enhance mathematical interpretability of the KB. \\
    \cline{2-4}

    & Knowledge Graphs 
    & \cite{CognitiveSemComKG, KnowledgeEnhancedReceiver} 
    & Store knowledge as entity-relation triples to support strong reasoning capabilities for error correction. \\
    \hline

    \multirow{5}{2.2cm}{\centering Mechanisms for Transmission} 
    & Residual Compression 
    & \cite{DeepLearningEmpoweredSharedKB} 
    & Transmit only residual information derived from KB fusion to significantly reduce symbol overhead. \\
    \cline{2-4}

    & Index-based Compression 
    & \cite{EndToEndGenerativeSKB} 
    & Transmit relevant indices to generate high-quality images based on class-level attributes. \\
    \cline{2-4}

    & GraphRAG-based Transmission 
    & \cite{fan2025kgrag} 
    & Combines GraphRAG with LLMs to extract minimal connected subgraphs for extreme compression and robust reconstruction. \\
    \cline{2-4}

    & Logic Error Correction 
    & \cite{CognitiveSemComKG} 
    & Use inference rules to detect and correct logically inconsistent triplets caused by channel noise. \\
    \cline{2-4}

    & Knowledge Extraction 
    & \cite{KnowledgeEnhancedReceiver} 
    & Design Transformer-based Knowledge Extractor to recover factual triples from noisy signals. \\
    \hline

    \multirow{3}{2.2cm}{\centering Deployment and Evolution} 
    & KB Update Mode 
    & \cite{ExplicitSemanticBase} 
    & Trigger update mechanism when communication intent changes or current KB fails to express new semantics. \\
    \cline{2-4}

    & AI-based Maintenance 
    & \cite{KnowledgeBaseEnabledGenerative} 
    & Highlight the role of generative AI in the automatic maintenance and updating of the KB. \\
    \cline{2-4}

    & Multi-user Distinction 
    & \cite{CognitiveSemComKG} 
    & Utilize user-specific private KGs and context information to distinguish and recover messages from different users. \\
    \hline
    \end{tabular*}
\end{table*}

\subsubsection{Architecture and Representation of Semantic KB}
To handle heterogeneous environments, KB construction exhibits hierarchical and modular traits. Ren et al. \cite{KnowledgeBaseEnabledGenerative} propose a generative architecture comprising source, task, and channel sub-bases to address cross-modal challenges. For multi-task scenarios, Wang et al. \cite{UnifiedHierarchicalSKB} introduce the unified hierarchical semantic KB, which employs horizontal and vertical constructions to maximize representation space and explore task correlations.
Regarding representation, methods include interpretable explicit Sebs \cite{ExplicitSemanticBase} and reasoning-capable Knowledge Graphs (KGs) \cite{CognitiveSemComKG, KnowledgeEnhancedReceiver} that store knowledge as entity-relation triples. Additionally, lightweight schemes utilizing class-level attribute vectors \cite{EndToEndGenerativeSKB} or shared textual databases \cite{DeepLearningEmpoweredSharedKB} have been developed to guide generative models and calculate semantic similarity.

\subsubsection{KB-Enabled Transmission}
\begin{figure}
    \centering
    \includegraphics[width=1\linewidth]{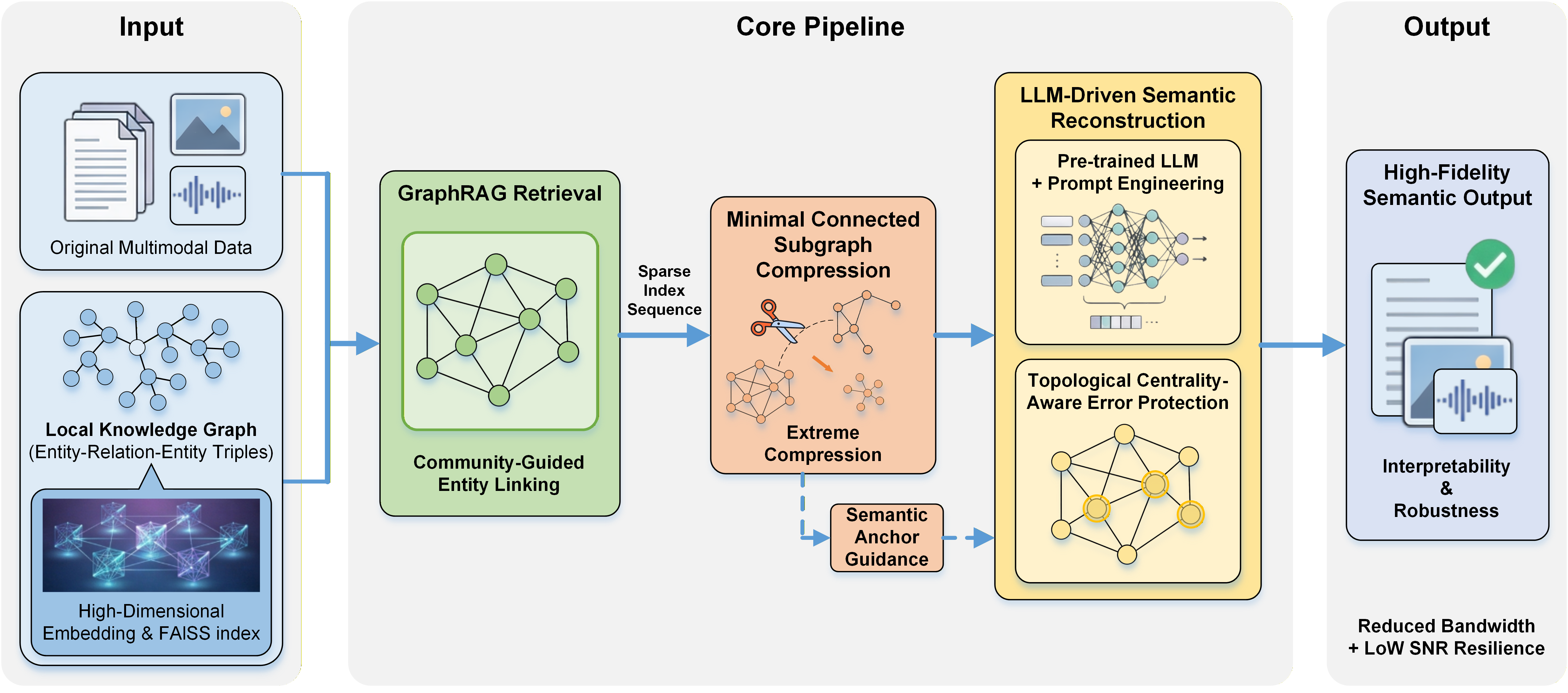}
    \caption{Illustration of the KGRAG-SC framework \cite{fan2025kgrag}, combining GraphRAG-guided semantic extraction, minimal connected subgraph compression, and knowledge-driven LLM reconstruction for robust and efficient semantic communication.}
    \label{fig:KG}
\end{figure}

Integrating KBs transforms transmission by enabling semantic compression and reasoning-based correction. For compression, shared KBs allow transmitting only residual information \cite{DeepLearningEmpoweredSharedKB} or relevant indices \cite{EndToEndGenerativeSKB} rather than complete data, significantly reducing overhead. 
More recently, retrieval-augmented generation has been introduced to optimize this process. As illustrated in Fig. \ref{fig:KG}, Fan et al. \cite{fan2025kgrag} propose the KGRAG-SC framework, which utilizes GraphRAG to extract minimal connected subgraphs. By transmitting only the compact node indices, it achieves extreme compression, while the receiver leverages the shared KB and LLMs to perform knowledge-driven semantic reconstruction. In terms of error correction, KGs provide prior knowledge for rectifying logically inconsistent triplets \cite{CognitiveSemComKG} or extracting factual triples from noisy signals \cite{KnowledgeEnhancedReceiver}. This reasoning capability transforms the receiver into a generative apprentice capable of high-fidelity reconstruction with minimal input \cite{LessDataMoreKnowledge}.

\subsubsection{KB Deployment and Evolution}
To adapt to dynamic environments, KBs require update mechanisms triggered by intent changes \cite{ExplicitSemanticBase} or automated maintenance via generative AI \cite{KnowledgeBaseEnabledGenerative}. Furthermore, distinct user contexts can be managed through private KGs \cite{CognitiveSemComKG}, while hierarchical designs enable single KBs to serve multiple downstream tasks efficiently \cite{UnifiedHierarchicalSKB}.

\subsection{Age of Semantic Information (AoSI)}
With the rapid growth of real-time applications, the Age of Information (AoI) has become a core metric for evaluating information freshness. AoI was originally derived from queueing theory to address the limitations of conventional throughput and delay metrics\cite{kaul2012real}.
Traditional RAN typically employs fixed or periodic update policies to maximize throughput or minimize end-to-end delay \cite{kaul2012real}. However, in SemRAN, not all updates carry equal value, and their utility depends on the query instant and the degree of information content change at the receiver \cite{delfani2024semantics}. Blindly increasing the update rate leads to network congestion and ineffective forwarding \cite{kaul2012real,chiariotti2022query}. Therefore, SemRAN should introduce semantic-aware update and scheduling mechanisms that generate and forward only truly meaningful updates.

In semantic-aware freshness analysis, three key semantic indicators are commonly considered. 
Semantic correctness evaluates whether the information is error-free and semantically consistent, penalizing misleading or incorrect updates \cite{li2024toward1}.
Semantic value reflects whether the delivered information is actually useful at the receiver, capturing its task-dependent utility at the moment of query or decision-making \cite{chiariotti2022query}. 
Semantic relevance measures how accurately the received information represents the current state of the source, indicating whether the content remains meaningful rather than merely outdated \cite{delfani2024semantics}. 
As illustrated in Table \ref{tab:semantic-aoi}, the semantic variants of AoI, AoSI, incorporate semantic value, relevance, and correctness, thereby providing finer-grained resource allocation and information dissemination for SemRAN.

\begin{table}[t]
\centering
\caption{Representative AoI Variants in SemRAN}
\label{tab:semantic-aoi}
\renewcommand{\arraystretch}{1.3}
\begin{tabular}{
    >{\centering\arraybackslash}m{2.0cm}
    >{\centering\arraybackslash}m{1.5cm}
    >{\centering\arraybackslash}m{1.5cm}
    >{\centering\arraybackslash}m{1.5cm}
}
\toprule
\textbf{AoI Variant} & \textbf{Sem. Corr.} & \textbf{Sem. Val.} & \textbf{Sem. Rel.} \\
\midrule
AoII\cite{maatouk2020age}  & \checkmark & - & - \\
QAoI\cite{chiariotti2022query}  & - & \checkmark & - \\
VAoI\cite{yates2021age}  & - & - & \checkmark \\
QVAoI\cite{delfani2024semantics} & - & \checkmark & \checkmark \\
\bottomrule
\end{tabular}

\vspace{2mm}
\footnotesize{
\textbf{Sem. Val.}: Semantic Value;\;  
\textbf{Sem. Rel.}: Semantic Relevance;\\
\textbf{Sem. Corr.}: Semantic Correctness.
}
\end{table}





\subsubsection{Age of Incorrect Information (AoII)} Maatouk \textit{et al.}\cite{maatouk2020age} propose AoII, which overcomes the timeliness-only limitation of conventional AoI by quantifying the cumulative penalty associated with incorrect information, thereby providing a more accurate assessment of both the effectiveness and relevance of status updates. This enables intelligent guidance of RAN resource allocation to prioritize the transmission of high-value semantic fragments. Li \textit{et al.} \cite{li2024toward1} explicitly incorporate AoII with other AoI variants into the proposed goal-oriented tensor unified framework, to support flexible and fine-grained goal-oriented SemCom.

\subsubsection{Query AoI (QAoI)} 
QAoI computes information age only when a query is generated at the receiver, ignoring non-query periods, which can effectively reduce unnecessary multi-hop forwarding and network overhead. This capability supports SemRAN in resource-constrained environments, by enabling dynamic scheduling and transmission prioritization of semantic fragments, thereby improving both the semantic effectiveness of query-driven decision-making and overall network energy efficiency \cite{chiariotti2022query}.

\subsubsection{Version AoI (VAoI)} 
Yates \cite{yates2021age} proposes VAoI, which measures information staleness in terms of version lag rather than absolute time, thereby eliminating the requirement for network-wide clock synchronization. It enables the network to accurately differentiate between ``merely outdated information” and ``information that has significantly diverged from the current true state and become semantically irrelevant,” thereby guiding resource scheduling to prioritize the transmission of high-value version updates that are both fresh and highly semantically relevant. 
Delfani \textit{et al.} \cite{delfani2024semantics} show that policies optimized for VAoI, compared to conventional AoI-based policies, can achieve comparable or superior semantic relevance and freshness while reducing unnecessary transmissions, thus enhancing both energy efficiency and decision-making quality in SemRAN.

\subsubsection{Query VAoI (QVAoI)} 
Delfani et al. \cite{delfani2024semantics} introduce QVAoI to jointly account for information freshness, relevance, and value under energy-constrained and unreliable channels. QVAoI integrates the characteristics of QAoI and VAoI, and evaluates the version lag only when a query is generated at the receiver, enabling semantic-aware transmission optimization. QVAoI allows SemRAN to deliver more timely, relevant, and valuable updates under the same energy budget, or to significantly reduce transmission frequency while maintaining high semantic quality. 


\section{Security Plane Techniques of SemRAN}
\label{section6}

\subsection{Security and Privacy Threat Issues}
\label{sectionsecurity}
The security and privacy threats faced by SemRAN are multi-level and full process. On the one hand, attackers no longer only focus on the stolen data itself, but also emphasize exploring the meanings carried by the stolen data, they targetedly steal or attack semantic information with key value \cite{guo2024survey}. On the other hand, considering that most semantic information in SemRAN is extracted and generated by DL models, attackers will not only launch attacks on the semantic transmission process, but also target DL models \cite{shen2023secure}. In this way, they undermine the accuracy and integrity of semantic information, posing severe challenges to the safe and reliable operation of SemRAN \cite{yang2024secure}. The security and privacy threat issues are outlined in Table \ref{securityissue}.

\begin{table*}[htbp] 
    \centering
    \caption{Security and Privacy Threat Issues in SemRAN}
    \label{securityissue}
    \resizebox{\textwidth}{!}{
        \begin{tabular}{cccccccccccccc}
            \toprule
            \multirow{2}{*}{Attack} & \multirow{2}{*}{Description} & \multicolumn{2}{c}{Target} & \multicolumn{3}{c}{Belonging layer} & \multicolumn{7}{c}{Compromised Properties} \\
            \cmidrule(lr){3-4} \cmidrule(lr){5-7} \cmidrule(lr){8-14}
            
             & & Trans. & Model & Phy. & Net. & Data link & Confidentiality & Integrity & Authentication & Access Control & Availability & Privacy & Freshness Secrecy \\
            \midrule
            
            Poisoning Attack & Inject malicious data during the training phase & - & \checkmark & \checkmark & \checkmark & - & - & \checkmark & - & - & \checkmark & - & \checkmark \\
            Backdoor Attack & Implant malicious triggers during the training phase & - & \checkmark & \checkmark & \checkmark & - & - & \checkmark & - & - & - & - & \checkmark \\
            Gradient Leakage & Overly transparent information sharing  & - & \checkmark & \checkmark & \checkmark & \checkmark & - & - & - & \checkmark & - & \checkmark & \checkmark \\
            Model Tampering & Manipulate model parameters or structure & - & \checkmark & \checkmark & \checkmark & \checkmark & \checkmark & \checkmark & - & \checkmark & \checkmark & - & \checkmark \\
            Semantic Noise & Transmission or encoding interference corrupts original semantics. & - & \checkmark & - & - & \checkmark & - & \checkmark & - & - & - & - & - \\
            Unaligned Knowledge Base & Mismatch in knowledge between parties & - & \checkmark & \checkmark & \checkmark & - & - & - & - & - & - & \checkmark & - \\
            Resource Bottlenecks & High model complexity or limited bandwidth & - & \checkmark & - & \checkmark & \checkmark & - & - & - & - & \checkmark & - & \checkmark \\
            Semantic Eavesdropping & Capture signals to steal key semantic information. & \checkmark & - & \checkmark & - & \checkmark & \checkmark & - & - & - & - & \checkmark & - \\
            Semantic Jamming & Disrupt the data stream or suppress energy & \checkmark & - & \checkmark & - & \checkmark & - & \checkmark & - & - & \checkmark & - & - \\
            Man-in-the-Middle & Intercept, tamper with, or forge signals & \checkmark & - & - & \checkmark & \checkmark & \checkmark & \checkmark & \checkmark & - & - & \checkmark & - \\
            Semantic Adversarial & Add small perturbations & \checkmark & - & \checkmark & - & - & - & \checkmark & - & - & \checkmark & - & - \\
            Semantic Inference & Analyze intermediate model information to infer sensitive data & \checkmark & - & \checkmark & \checkmark & - & - & - & - & - & - & \checkmark & - \\
            Model Inversion & Eavesdrop on transmitted data to reconstruct input information. & \checkmark & - & \checkmark & \checkmark & - & - & - & - & - & - & \checkmark & - \\

            \bottomrule
        \end{tabular}
    }
\end{table*}

\subsubsection{Attacks Against Semantic Models}
During the training and deployment of semantic models, the following security threats may occur.
\begin{itemize}
\item 
\textbf{Poisoning Attack:}
It impairs the performance stability and output integrity of target models by injecting malicious or misleading data during the training phase\cite{peng2024adversarial}. 
\item 
\textbf{Backdoor Attack:}
It implants malicious data into some training samples. After being activated by specific conditions during the inference phase, it induces the compromised model to output incorrect results only for target input samples\cite{sagduyu2023vulnerabilities}. 
\item 
\textbf{Gradient Leakage:}
Attackers steal exposed shared gradient information and infer training data, labels, and even sensitive semantic information in reverse\cite{yang2023gradient}.
\item 
\textbf{Model Tampering Attack:}
It manipulates the parameters or structure of semantic models, causing them to output incorrect semantics, leak sensitive information, or lose semantic transmission function \cite{meng2025survey}.
\item 
\textbf{Resource Bottleneck:}
In the federated training semantic models, information bottleneck attacks caused by issues such as high model complexity, limited bandwidth, and device heterogeneity will lead to increased communication overhead, latency, and disconnection, and may even result in training disruption\cite{meng2025survey}.
\item 
\textbf{Semantic Noise:}
It causes semantic information to be misunderstood, leading to decoding errors and causing deviations in the transmission and reception of semantic symbols. It also manifests differently in different data types. Text is semantically ambiguous such as word replacement, and images are filled with adversarial samples that are difficult for humans to detect\cite{hu2022robust}.
\item 
\textbf{Unaligned Knowledge Base:}
It causes interruption of semantic reasoning paths and distortion of semantic decoders, which impairs the reliability of KB systems \cite{lan2021survey}.

\end{itemize}

\subsubsection{Attacks Against Semantic Transmission}
In the semantic transmission process, it mainly faces the following threats:
\begin{itemize}
\item 
\textbf{Semantic Eavesdropping Attack:}
It can exploit the openness of channels and use specialized receiving devices to capture and parse transmitted signals, thereby stealing key semantic information \cite{meng2025survey}.
\item 
\textbf{Semantic Jamming Attack:}
It disrupts the transmitted data stream or implements energy suppression, reducing the consistency between the decoded results and original semantics, and ultimately interfering with the receiver's accurate understanding of semantic information\cite{tang2023gan}.
\item 
\textbf{Man-in-the-Middle Semantic Attack:}
It involves intercepting legitimate communication signals, then either tampering with the signal content before forwarding it to the receiver or directly sending forged semantic information to the receiver \cite{meng2025generative}. Meanwhile, attackers can steal or infer users’ sensitive privacy information based on the intercepted messages\cite{shen2023secure}.
\item 
\textbf{Semantic Adversarial Attack:}
It adds small and imperceptible perturbations to the input signal to induce SemRAN to generate incorrect semantic representations, thereby undermining the accuracy of semantic information processing\cite{do2025security}.
\item 
\textbf{Semantic Inference Attack:}
It is mainly divided into three sub-categories: membership inference attacks, property inference attacks, and reconstruction attacks. They all analyze the intermediate information exposed during the model training process to determine whether specific data belongs to the training set members, mine sensitive attributes in the personal knowledge base, and reconstruct sensitive semantic information\cite{meng2025survey}.
\item 
\textbf{Model Inversion Attack:}
It infers or reconstructs sensitive input information by eavesdropped data\cite{wang2025diffusion}.

\end{itemize}

\subsection{Security and Privacy Protection Techniques}

As illustrated in Table \ref{securityprotection}, to counter the complex security threats faced by SemRAN, researchers have proposed various defense strategies, which are described in detail as follows.

\begin{table*}
\caption{Security and Privacy Protection Techniques for SemRAN}
\label{securityprotection}
\centering
\renewcommand{\arraystretch}{1.4} 
\begin{tabular}{|>{\centering\arraybackslash}m{1.8cm}|>{\centering\arraybackslash}m{4cm}|>{\centering\arraybackslash}m{7cm}|>{\centering\arraybackslash}m{2cm}|}
\hline
\multicolumn{2}{|c|}{\textbf{Defense Methods}} & \multicolumn{1}{c|}{\textbf{Defensive Attacks}} & \textbf{Representative Schemes} \\ \hline

\multicolumn{2}{|c|}{Data Cleaning} & Poisoning Attacks, Backdoor Attacks & \cite{rajan2024data} \\ \hline

\multicolumn{2}{|c|}{Robust Enhancing} & Poisoning Attacks, Backdoor Attacks & \cite{raha2025advancing} \\ \hline

\multicolumn{2}{|c|}{Differential Privacy} & Gradient Leakage, Semantic Inference Attacks & \cite{chen2024enhancing,liu2024adaptive}\\ \hline

\multicolumn{2}{|c|}{Model Compression} 
& Poisoning Attacks, Backdoor Attacks, Model Inversion Attacks, Resource Bottlenecks & \cite{zhou2024backdoor} \\ \hline

\multicolumn{2}{|c|}{Adversarial Training} & Poisoning Attacks, Backdoor Attacks, Semantic Noise, Semantic Adversarial Attacks &\cite{liu2023semprotector,peng2024adversarial,kang2023adversarial,zhou2025rome}  \\ \hline

\multicolumn{2}{|c|}{Blockchain} & Unaligned Knowledge Base & \cite{lin2025blockchain}  \\ \hline

\multirow{5}{=}{Cryptography Technology} & Homomorphic Encryption & Semantic Eavesdropping Attacks & \cite{meng2025secure,yuan2025homomorphic} \\ \cline{2-4} 
 & Secure Multi-party Computation & Semantic Inference Attacks &  / \\ \cline{2-4} 
 & Trusted Execution Environment & Semantic Inference Attacks & \cite{zhang2025balancing} \\ \cline{2-4} 
 & Secure Aggregation & Resource Bottlenecks &  / \\ \cline{2-4} 
 & Quantum Cryptography & Semantic Eavesdropping Attacks, Man-in-the-Middle Semantic Attacks &\cite{rizvi2025controlled}\\ \hline

\multicolumn{2}{|c|}{Semantic Steganography Communication} & Semantic Eavesdropping Attacks & \cite{wang2025image,gao2025semstediff}\\ \hline

\multicolumn{2}{|c|}{Covert Semantic Communication} & Semantic Eavesdropping Attacks, Semantic Jamming Attacks & \cite{liu2025learning,zhang2025optimization}\\ \hline

\multirow{5}{=}{Physical-layer Security} & Beamforming & Semantic Eavesdropping Attacks, Semantic Jamming Attacks &\cite{dai2024secure,zhousecurity} \\ \cline{2-4} 
 & Artificial Noise & Semantic Eavesdropping Attacks &\cite{tan2024security} \\ \cline{2-4} 
 & Relay Cooperation & Man-in-the-Middle Semantic Attacks & \cite{tang2023cooperative} \\ \cline{2-4} 
 & Physical-layer Key Generation & Semantic Eavesdropping Attacks & \cite{zhao2022semkey} \\ \cline{2-4} 
 & Physical-layer Authentication & Man-in-the-Middle Semantic Attacks & \cite{gao2023esanet} \\ \hline
\end{tabular}
\end{table*}

\begin{itemize}
\item 
\textbf{Data Cleaning:}
It ensures the accuracy of transmitted data by filtering noise, removing malicious information, and removing abnormal and redundant data\cite{chu2016data}.
\item 
\textbf{Robust Enhancing:}
It enhances SemRAN's ability to resist interference such as noise, malicious attacks, and channel fluctuations\cite{raha2025advancing}.
\item 
\textbf{Differential Privacy:}
It desensitizes dataset computation results to a single entry, realizing privacy protection \cite{liu2023adaptable}. 
\item 
\textbf{Model Compression:} It can effectively eliminate parameter redundancy, save storage space, and reduce communication costs, thus strengthening the defense of SemRAN against multiple attacks. It mainly includes the following techniques. Firstly, model pruning optimizes performance or eliminates security risks by removing redundant or threat related parameters from semantic models\cite{he2023structured}. Secondly, parameter quantization converts high-precision weights in semantic models into low precision formats, significantly reducing model volume, computation, and storage overhead while minimizing model performance\cite{park2024vision}.
Thirdly, low-rank decomposition simplifies the semantic model structure and compresses redundant information\cite{swaminathan2020sparse}.
Fourthly, knowledge distillation ensures semantic model's security by having multiple teacher models split and process sensitive data or leveraging the robustness of teacher models to resist adversarial perturbations \cite{gou2021knowledge}.
\item 
\textbf{Adversarial training:}
It generates adversarial samples with small perturbations or introduces adversarial mechanisms to train semantic models in scenarios with interference, thereby achieving resistance to attacks \cite{peng2024adversarial}.
\item 
\textbf{Blockchain:}
It utilizes decentralized storage, encrypted transmission, and tamper proof features, to protect the integrity and confidentiality of semantic information \cite{lin2025blockchain}.
\item 
\textbf{Cryptography Technology:}
Emerging cryptography technologies provide a balance between data protection and semantic processing. For example, the ciphertext generated by homomorphic encryption preserves semantic features, which can be extracted by semantic models \cite{meng2025secure,yuan2025homomorphic}.
Besides, secure multi-party computation enables multiple parties to collaboratively perform computations without disclosing their private data \cite{zhou2024secure}. Additionally, trusted execution environment provides a secure execution space for sensitive code and data independent of untrusted software through hardware-assisted isolation mechanisms, memory encryption, and code integrity verification, resisting unauthorized access and various attacks \cite{jauernig2020trusted}.
In addition, secure aggregation achieves the aggregated computation of client model updates in federated learning while resisting semi-honest or malicious attackers to protect data and model privacy \cite{zhou2022survey}. Furthermore, quantum cryptography relies on the fundamental properties of quantum mechanics, such as superposition and entanglement, to secure SemRAN \cite{meng2025survey}.
\item 
\textbf{Semantic Steganography Communication:} It embeds private images into the cover image, while semantic eavesdroppers can only decode the steganographic image to obtain the cover image \cite{wang2025image,gao2025semstediff}.
\item 
\textbf{Covert Semantic Communication:}
It hides signal power under wireless channel noise, making it difficult for eavesdroppers to detect the communication behavior itself, thus achieving imperceptibility in SemRAN \cite{liu2025learning}.
\item 
\textbf{Physical Layer Security:} Unlike cryptography techniques, physical layer security is regarded as endogenous security technologies in SemRAN, as it leverages the unique characteristics of wireless channels such as randomness, time-varying nature, and unforgeability \cite{meng2025survey2}.
For example, beamforming generates beams with specific directionality by adjusting the phase and amplitude of the signals, thereby significantly reducing the signal quality received by eavesdroppers \cite{meng2025survey}. Besides, artificial noise reduces the received SNR of eavesdroppers, thereby ensuring information security \cite{tan2024security}. Additionally, relay cooperation ensures security by leveraging trusted relays for forwarding, thus improving secrecy rate \cite{rodriguez2015physical}.
In addition, the complex and random channel environment in SemRAN allows both parties to capture highly correlated and spatiotemporal physical layer key, thereby enhancing the anti-eavesdropping ability \cite{jin2022ris}. Furthermore, physical layer authentication ensures communication security by exploiting inherent physical features of wireless devices \cite{meng2022multiuser}, such as radio frequency fingerprints \cite{meng2023multidimensional} and channel fingerprints \cite{meng2022physical}, to achieve device identity verification and attack detection in SemRAN \cite{gao2023esanet}.
\end{itemize}

\section{Applications and Standardization of SemRAN}
\label{section7}

\subsection{SemRAN for New Services}
SemRAN can support the following new services.

\begin{itemize}
\item 
\textbf{HRLLC:}
SemRAN can improve resource utilization efficiency and effectively reduce bandwidth requirements, with low requirements for channel quality and strong anti-interference ability, thus satisfying the requirements of HRLLC for future 6G networks \cite{zhang2025flexible}. 
\item 
\textbf{AIGC:}
AIGC can perform intelligent content generation, data compression, and cache optimization at the network edge, improving the overall efficiency of mobile communications. SemRAN can enhance the performance of AIGC by assisting its deployment with its efficient comprehension. Du \textit{et al.} \cite{du2024enabling} employ SemRAN to accurately capture user needs, to facilitate the dynamic and optimal selection of AIGC service providers.
\item 
\textbf{ISCCI:}
ISCCI can comprehensively perceive its environment, process data, and communicate, fully leveraging the network's potential with intelligent capabilities. SemRAN, through its semantic propagation-based collaborative framework, more accurately responds to network demands and provides support for ISCCI \cite{zhu2023semantic}. 
\end{itemize}


\subsection{SemRAN for New Applications}
SemRAN can support the following new applications.
\begin{itemize}
\item 
\textbf{Immersive Communications:}
Immersive Communications delivers highly immersive interactive experiences for users in the future infinite network through technologies like augmented reality (AR), virtual reality (VR), and extended reality (XR). SemRAN dramatically reduces the data volume to be transmitted, significantly conserving bandwidth for immersive communications. For example, Wang \textit{et al.} \cite{wang2023adaptive} employ SemRAN to improve transmission quality and higher equivalent semantic rate with limited resources of Immersive Communications.
\item 
\textbf{Agent-to-Agent Communications:}
Agent-to-Agent communications empower network elements to autonomously execute self-optimization, self-healing, and self-configuration, thereby substantially diminishing operational expenditures while augmenting network resilience for forthcoming complex network environments. SemRAN enhances these agents with advanced cognitive and communicative capabilities, thereby reducing signaling overhead and accelerating decision-making processes \cite{zhou2024semantic}. 
\item 
\textbf{SAGSIN:}
SAGSIN integrates space, air, ground, and sea communication to deliver seamless global coverage capabilities, which aims to solve the challenges in areas with poor communication environment, such as forests, deserts, oceans, and disaster areas. SemRAN delivers solutions for challenging environments where SAGSIN struggles, such as large-scale scenarios, highly dynamic channels, and limited device capabilities \cite{cao2025channel,meng2024semantics,bian2025multi}.\item 
\textbf{Intelligent Healthcare:}
Intelligent healthcare utilizes seamless inter-device data exchange and efficient data analytics to deliver smart medical services. SemRAN can address critical multi-modal challenges, including diverse data formats and cross-system interoperability, while ensuring consistent semantic interpretation \cite{balakrishnan2025ontology}. 
\item 
\textbf{Smart Factory:}
Smart factory requires a high-efficiency, low-latency communication network to support data interaction and collaborative operation among a large number of devices. SemRAN can reduce the transmission of redundant information and accelerate decision-making by deploying semantic-aware edge computing nodes and semantic-level communication optimization, realizing more agile factory response \cite{hu2021manufacturing}. 
\item 
\textbf{Intelligent Transportation:}
As a critical solution addressing modern urban transportation demands for efficiency, safety, and intelligence, intelligent transportation necessitate high-bandwidth, efficient, and accurate data transmission. SemRAN, with its superior anti-interference capabilities, ensures enhanced accuracy and efficiency in information transmission to support intelligent transportation \cite{wang2024polling}. 
\end{itemize}

\begin{table}
\centering
\caption{Overview of Focused Directions of IMT-2030 (6G) Semantic Communication Task Group.}
\label{tab:imt2030_semcom}
\renewcommand{\arraystretch}{1.2} 
\begin{tabular}{|p{1.4cm}|p{6cm}|}
\hline
\textbf{Company} & \textbf{Focus Direction} \\ \hline
BUPT & Typical scenario, performance index, simulation platform and verification system, cross-layer design, JSCC, MDMA, HARQ, semantic information theory, semantic synchronization, NTSCC, entropy-based importance resampling, FDD, TDD, E2E design, environmental semantics, privacy protection, encrypt system, CSI feedback, agent-to-agent communication, DVST, full-motion video transmission, model management, code book, diffusion-based image transmission. \\ \hline
CMCC & Typical scenario, performance index, HARQ. \\ \hline
OPPO & JSCC, CSI feedback, control semantics. \\ \hline
ZTE & Semantic channel process. \\ \hline
PCL & Typical scenario, performance index, simulation platform and verification system, semantic KB, cloud-edge-device model collaboration, generative SemCom, agent-to-agent communication. \\ \hline
SJTU & Semantic codebook, broadcasting, digital SemCom, MambaJSCC, bit-level JSCC. \\ \hline
Huawei & Typical scenario, M2M SemCom, LLM-based SemCom. \\ \hline
Inspur & Typical scenario, performance index, JSCC. \\ \hline
BIT & JSCC, CSI feedback. \\ \hline
VIVO & Cross-layer design, typical scenario, data check. \\ \hline
ZGC-NET & Typical scenario, channel estimation, agent-to-agent communication, video transmission, diffusion-based image transmission, generative SemCom, JSCC, code book, E2E design, model management. \\ \hline
THU & Task-oriented SemCom, multi-model semantic, satellite network, computility network-based SemCom. \\ \hline
UESTC & FDD, partial sampling-based SemCom. \\ \hline
BIT & Speech SemCom, semantic encoding for MIMO, multiuser SemCom. \\ \hline
Xiaomi & JSCC, CSI feedback. \\ \hline
ZJU & Generative SemCom. \\ \hline
Lenovo & QoS. \\ \hline
Nokia & Token communication. \\ \hline
\end{tabular}
\end{table}

\subsection{Standardization Progress of SemRAN}


\begin{itemize}
\item \textbf{CCSA:} 
The ``Research on Intellicise Communication Technologies'' Project was initiated by the technology working committee on wireless communication (TC5) of CCSA in 2022 and has received participation and support from multiple operators, equipment vendors, and university research institutes. It releases the research report on ``Key Technologies of Intellicise Communication", systematically elaborating on innovative intellicise communication and related key technology research for future 6G. It explores the evolution path of 6G, including main driving forces, application scenarios, key technologies, simulation platforms, and verification systems of intellicise communication. It aims to improve the transmission capacity, spectrum efficiency, and business service provision capabilities of 6G. Furthermore, on June 10, 2025, the first plenary meeting of the standard promotion committee on technologies and standardizations of SemCom and intellicise communication (TC630) of CCSA was successfully held at BUPT. The promotion committee will focus on the concept condensation, standard research, and industry collaboration of SemCom and intellicise communication, and promote the construction of technical systems and verification of key capabilities.
\item \textbf{IMT-2030:} The IMT-2030 wireless technology working group establishes the IMT-2030 SemCom task group, responsible for researching key wireless SemCom technologies and standardization work for 6G. As illustrated in Table \ref{tab:imt2030_semcom}, the SemCom standardization direction gradually refined involves multiple aspects, such as typical scenarios, performance index, JSCC, CSI feedback, MDMA, semantic KB, etc., and continues to promote consensus on SemCom standardization for 6G.
\item \textbf{3GPP:} Table \ref{3gpp_proposals} provides some representative 3GPP proposals for SemRAN. Notably, the proposal for 6G scenario use case of SemCom, ``use case on AI-driven multi-vehicle cooperative perception", was successfully accepted at the 110th Fukuoka Conference of 3GPP SA1, laying an important foundation for the standardization of key SemRAN technologies.
\end{itemize}

\begin{table*}[t]
	\centering
	\caption{Overview of Representative 3GPP Proposals for SemRAN.}
	\label{3gpp_proposals}
	\renewcommand{\arraystretch}{1.4} 
	
	\begin{tabular}{|
	>{\raggedright\arraybackslash}m{2.4cm}|
	>{\raggedright\arraybackslash}m{1.3cm}|
	>{\raggedright\arraybackslash}m{2.8cm}|
	>{\raggedright\arraybackslash}m{8.3cm}|}
	\hline
	\textbf{Meeting} & \textbf{Location} & \textbf{Time} & \textbf{Title} \\ \hline
		
	\multirow{7}{=}{3GPP TSG-SA WG1 Meeting} 
	& \multirow{5}{=}{Fukuoka, Japan} 
	& \multirow{5}{=}{19--23 May, 2025} 
	& Use case on AI-enabled low-altitude UAV inspection \\ \cline{4-4} 
	
	& & & Use case on AI-driven satellite remote sensing and transmission \\ \cline{4-4} 
	& & & Use case on 6G multiple AI agent collaboration \\ \cline{4-4} 
	& & & Use case on AI-driven multi-vehicle cooperative perception \\ \cline{2-4} 
	
	& \multirow{2}{=}{Goteborg, Sweden} 
	& \multirow{2}{=}{25--29 August, 2025} 
	& Use case on 6G AI Agents collaboration for disaster rescue \\ \cline{4-4} 
	
	& & & Use case on AI-enabled satellite-UAV collaborative emergency service \\ \cline{4-4} 
	& & & Use case on 3D hyper-realistic video services \\ \hline
		
	3GPP TSG-RAN WG1 Meeting 
	& Bengaluru, India 
	& 25--29 August, 2025 
	& Discussion on AI/ML-enabled use cases for 6GR \\ \hline
		
	3GPP TSG-RAN WG Meeting 
	& Beijing, China 
	& 15--18 September, 2025 
	& Requirements on the Integration of AI and 6G RAN \\ \hline
		
	\multirow{2}{=}{3GPP TSG-RAN WG1 Meeting} 
	& \multirow{2}{=}{Prague, Czech Republic} 
	& \multirow{2}{=}{13--17 October, 2025} 
	& Discussion on AI/ML-enabled use cases for 6GR \\ \cline{4-4} 
	
	& & & Views on inter-vendor training collaboration for two-sided AI/ML models \\ \hline
		
	\end{tabular}
\end{table*}

\section{Challenges and Future Directions}
\label{section8}

\subsection{Theoretical Framework of SemRAN}
Classical information theory, which is built upon bit-level representations and predefined codebooks, is insufficient to characterize semantic information, semantic distortion, and task-oriented performance. In SemRAN, semantic relevance, semantic correctness, and task utility are closely coupled with shared knowledge and inference processes, which cannot be adequately captured by conventional metrics. Without a matched theoretical foundation, it remains difficult to formally analyze semantic efficiency, optimize system design, or provide performance guarantees. Niu and Zhang \cite{niu2025mathematical} propose semantic information theory and prove that SemCom can significantly improve the spectral efficiency of communication systems based on synonymous mapping. In the future, more theoretical research is needed to guide the development of SemRAN.

\subsection{Management of Semantic Models in SemRAN}
SemRAN is expected to support diverse semantic models tailored to different tasks, data modalities, and service requirements. However, managing these models in a unified SemRAN remains challenging. Issues such as semantic model selection, model switching, and cross-model compatibility are not yet well addressed. Achieving harmonized orchestration of semantic models is therefore an important research direction for scalable and flexible SemRAN deployments.

\subsection{Hybrid Semantic-Bit Coexisting Communication}
In practical RANs, SemCom will coexist with conventional bit-based services rather than fully replacing them \cite{zhang2025beamforming}. This coexistence introduces challenges in the design of SemRAN, as the two communication paradigms pursue different objectives and rely on incompatible performance indicators. Coordinating semantic-oriented and bit-oriented transmissions within SemRAN requires a unified framework for performance evaluation and protocol interaction. Moreover, ensuring stable end-to-end behavior in hybrid transmission paths remains an open issue for SemRAN.

\subsection{Security and Privacy Protection of SemRAN}
As analyzed in Section \ref{sectionsecurity}, SemCom brings significant benefits to RAN but also introduces new security and privacy challenges that do not exist in conventional RAN. 
Although researchers have proposed many defense methods to enhance the security of SemRAN, there is still a gap in research on defense methods for some attacks, such as the model tampering attacks. 
The current authentication schemes mostly focus on user identity and do not consider the identity of semantic models. Attackers can interfere with the semantic encoding and decoding performance of legitimate users by forging the identity of semantic models, and this problem urgently needs to be solved.


\section{Conclusions}
\label{section9}
In this survey, we first introduced the concept of RAN and SemCom. Subsequently, we analyzed what RAN can benefit from SemCom, and proposed the architecture of SemRAN, which is composed of physical layer, data link layer, network layer, and security plane. Furthermore, we comprehensively surveyed existing SemRAN techniques, covering semantic-based coding, semantic-aware CSI feedback, semantic-based beam management, and semantic-based MA in the physical layer; semantic-based HARQ and resource scheduling in the data link layer; semantic-based KB and AoSI in the network layer; and security threats and defense technologies in the security plane. Moreover, we envisioned the future services and applications, and outlined current standardization efforts for SemRAN. Ultimately, we highlighted existing challenges and provided directional guidance for future research in SemRAN.


\bibliography{ref.bib}

@article{wp5d2022future,
  title={Future technology trends of terrestrial international mobile telecommunications systems towards 2030 and beyond},
  author={WP5D, ITUR},
  journal={International Telecommunication Union, Report M},
  pages={2516--0},
  year={2022}
}

@article{qin2023review,
  title={A review of codebooks for CSI feedback in 5G new radio and beyond},
  author={Qin, Ziao and others},
  journal={arXiv preprint arXiv:2302.09222},
  year={2023}
}

@article{yang2022semantic,
  title={Semantic communications for future internet: Fundamentals, applications, and challenges},
  author={Yang, Wanting and others},
  journal={IEEE Communications Surveys \& Tutorials},
  volume={25},
  number={1},
  pages={213--250},
  year={2023},
  publisher={IEEE}
}

@article{xu2025semantic,
  title={Semantic Prior Aided Channel-Adaptive Equalizing and De-Noising Semantic Communication System with Latent Diffusion Model},
  author={Xu, Bingxuan and others},
  journal={IEEE Transactions on Wireless Communications},
  year={2025},
  publisher={IEEE}
}

@article{gao2018compressive,
  title={Compressive sensing techniques for next-generation wireless communications},
  author={Gao, Zhen and others},
  journal={IEEE Wireless Communications},
  volume={25},
  number={3},
  pages={144--153},
  year={2018},
  publisher={IEEE}
}

@article{guo2024deep,
  title={Deep learning for CSI feedback: One-sided model and joint multi-module learning perspectives},
  author={Guo, Yiran and others},
  journal={IEEE Communications Magazine},
  year={2024},
  publisher={IEEE}
}

@article{parvez2018survey,
  title={A survey on low latency towards 5G: RAN, core network and caching solutions},
  author={Parvez, Imtiaz and others},
  journal={IEEE Communications Surveys \& Tutorials},
  volume={20},
  number={4},
  pages={3098--3130},
  year={2018},
  publisher={IEEE}
}

@article{gao2025adaptive,
  title={Adaptive Cross-Modal Super-Resolution Semantic Communication for Mobile AI-Generated Panoramic Video},
  author={Gao, Haixiao and others},
  journal={IEEE Transactions on Cognitive Communications and Networking},
  year={2025},
  publisher={IEEE}
}

@article{xu2023task,
  title={Task-oriented and semantic-aware heterogeneous networks for artificial intelligence of things: Performance analysis and optimization},
  author={Xu, Xiaodong and others},
  journal={IEEE Internet of Things Journal},
  volume={11},
  number={1},
  pages={228--242},
  year={2023},
  publisher={IEEE}
}

@article{wang2024feature,
  title={Feature importance-aware task-oriented semantic transmission and optimization},
  author={Wang, Yining and others},
  journal={IEEE Transactions on Cognitive Communications and Networking},
  volume={10},
  number={4},
  pages={1175--1189},
  year={2024},
  publisher={IEEE}
}

@article{jiang2025large,
  title={From large ai models to agentic ai: A tutorial on future intelligent communications},
  author={Jiang, Feibo and others},
  journal={arXiv preprint arXiv:2505.22311},
  year={2025}
}

@article{zhang2025resource,
  title={Resource allocation in wireless semantic communications: A comprehensive survey},
  author={Zhang, Chujun and others},
  journal={IEEE Communications Surveys \& Tutorials},
  year={2025},
  publisher={IEEE}
}

@article{brik2024explainable,
  title={Explainable ai in 6g o-ran: A tutorial and survey on architecture, use cases, challenges, and future research},
  author={Brik, Bouziane and others},
  journal={IEEE Communications Surveys \& Tutorials},
  year={2024},
  publisher={IEEE}
}

@article{chen2024evolution,
  title={Evolution of RAN architectures toward 6G: Motivation, development, and enabling technologies},
  author={Chen, Jiacheng and others},
  journal={IEEE Communications Surveys \& Tutorials},
  volume={26},
  number={3},
  pages={1950--1988},
  year={2024},
  publisher={IEEE}
}

@article{herrera2025tutorial,
  title={A Tutorial on O-RAN Deployment Solutions for 5G: From Simulation to Emulated and Real Testbeds},
  author={Herrera, Juan Luis and others},
  journal={IEEE Communications Surveys \& Tutorials},
  year={2025},
  publisher={IEEE}
}

@article{santos2025managing,
  title={Managing O-RAN networks: xApp development from zero to hero},
  author={Santos, Joao F and others},
  journal={IEEE Communications Surveys \& Tutorials},
  year={2025},
  publisher={IEEE}
}

@article{alam2025comprehensive,
  title={A comprehensive tutorial and survey of O-RAN: Exploring slicing-aware architecture, deployment options, use cases, and challenges},
  author={Alam, Khurshid and others},
  journal={IEEE Communications Surveys \& Tutorials},
  year={2025},
  publisher={IEEE}
}

@article{teng2025conquering,
  title={Conquering High Packet-Loss Erasure: MoE Swin Transformer-Based Video Semantic Communication},
  author={Teng, Lei and others},
  journal={arXiv preprint arXiv:2508.01205},
  year={2025}
}

@inproceedings{xu2023latent,
  title={Latent semantic diffusion-based channel adaptive de-noising SemCom for future 6G systems},
  author={Xu, Bingxuan and others},
  booktitle={GLOBECOM 2023-2023 IEEE Global Communications Conference},
  pages={1229--1234},
  year={2023},
  organization={IEEE}
}

@article{khan2023ai,
  title={AI-RAN in 6G Networks: State-of-the-Art and Challenges},
  author={Khan, Naveed Ali and Schmid, Stefan},
  journal={IEEE Open Journal of the Communications Society},
  volume={5},
  pages={294--311},
  year={2023},
  publisher={IEEE}
}

@article{zhang2025comai,
  title={ComAI: The Convergence of Communication and Artificial Intelligence},
  author={Zhang, Ping and others},
  journal={IEEE Communications Surveys \& Tutorials},
  year={2025},
  publisher={IEEE}
}

@article{meng2025survey,
  title={A survey of secure semantic communications},
  author={Meng, Rui and others},
  journal={Journal of Network and Computer Applications},
  pages={104181},
  year={2025},
  publisher={Elsevier}
}

@article{yining2024intellicise,
  title={Intellicise model transmission for semantic communication in intelligence-native 6G networks},
  author={Yining, Wang and others},
  volume={21},
  number={7},
  pages={95--112},
  year={2024},
  publisher={IEEE}
}

@article{guo2024survey,
  title={A survey on semantic communication networks: Architecture, security, and privacy},
  author={Guo, Shaolong and others},
  journal={IEEE Communications Surveys \& Tutorials},
  year={2024},
  publisher={IEEE}
}

@article{won2024resource,
  title={Resource management, security, and privacy issues in semantic communications: A survey},
  author={Won, Dongwook and others},
  journal={IEEE Communications Surveys \& Tutorials},
  volume={27},
  number={3},
  pages={1758--1797},
  year={2024},
  publisher={IEEE}
}

@article{sun2025s,
  title={S-RAN: Semantic-aware radio access networks},
  author={Sun, Yao and others},
  journal={IEEE Communications Magazine},
  year={2025},
  volume={63},
  number={4},
  pages={207-213},
  publisher={IEEE}
}

@article{lu2025important,
  title={Important bit prefix m-ary quadrature amplitude modulation for semantic communications},
  author={Lu, Haonan and others},
  journal={arXiv preprint arXiv:2508.11351},
  year={2025}
}

@article{wang2023road,
  title={On the road to 6G: Visions, requirements, key technologies, and testbeds},
  author={Wang, Cheng-Xiang and others},
  journal={IEEE Communications Surveys \& Tutorials},
  volume={25},
  number={2},
  pages={905--974},
  year={2023},
  publisher={IEEE}
}

@article{polese2023understanding,
  title={Understanding O-RAN: Architecture, interfaces, algorithms, security, and research challenges},
  author={Polese, Michele and others},
  journal={IEEE Communications Surveys \& Tutorials},
  volume={25},
  number={2},
  pages={1376--1411},
  year={2023},
  publisher={IEEE}
}

@article{shi2025band,
  title={In-Band Full-Duplex System for Semantic Communication},
  author={Shi, Mengran and others},
  journal={IEEE Internet of Things Journal},
  year={2025},
  publisher={IEEE}
}

@article{zhang2024intellicise,
  title={Intellicise wireless networks from semantic communications: A survey, research issues, and challenges},
  author={Zhang, Ping and others},
  journal={IEEE Communications Surveys \& Tutorials},
  year={2024},
  publisher={IEEE}
}

@article{ren2025semcsinet,
  title={SemCSINet: A Semantic-Aware CSI Feedback Network in Massive MIMO Systems},
  author={Ren, Ruonan and others},
  journal={arXiv preprint arXiv:2505.08314},
  year={2025}
}

@inproceedings{zheng2025semantic,
  title={Semantic diversity for massive MIMO CSI feedback},
  author={Zheng, Zhe and others},
  booktitle={2025 10th International Conference on Computer and Communication System (ICCCS)},
  pages={831--837},
  year={2025},
  organization={IEEE}
}

@article{zhang2024scan,
  title={SCAN: Semantic communication with adaptive channel feedback},
  author={Zhang, Guangyi and others},
  volume={10},
  number={5},
  pages={1759--1773},
  year={2024},
  publisher={IEEE}
}

@inproceedings{cao2023adaptive,
  title={Adaptive csi feedback with hidden semantic information transfer},
  author={Cao, Jiaqi and others},
  booktitle={ICASSP 2023-2023 IEEE International Conference on Acoustics, Speech and Signal Processing (ICASSP)},
  pages={1--5},
  year={2023},
  organization={IEEE}
}

@article{gao2023hybrid,
  title={Hybrid knowledge-data driven channel semantic acquisition and beamforming for cell-free massive MIMO},
  author={Gao, Zhen and others},
  journal={IEEE journal of selected topics in signal processing},
  volume={17},
  number={5},
  pages={964--979},
  year={2023},
  publisher={IEEE}
}

@article{zhu2024semantic,
  title={Semantic-based channel state information feedback for uav-assisted isac systems},
  author={Zhu, Guyue and others},
  journal={IEEE Internet of Things Journal},
  year={2024},
  publisher={IEEE}
}

@article{giordani2018tutorial,
  title={A tutorial on beam management for 3GPP NR at mmWave frequencies},
  author={Giordani, Marco and others},
  journal={IEEE Communications Surveys \& Tutorials},
  volume={21},
  number={1},
  pages={173--196},
  year={2018},
  publisher={IEEE}
}

@inproceedings{imran2023environment,
  title={Environment semantic aided communication: A real world demonstration for beam prediction},
  author={Imran, Shoaib and others},
  booktitle={2023 IEEE International Conference on Communications Workshops (ICC Workshops)},
  pages={48--53},
  year={2023},
  organization={IEEE}
}

@article{xue2024survey,
  title={A survey of beam management for mmWave and THz communications towards 6G},
  author={Xue, Qing and others},
  journal={IEEE Communications Surveys \& Tutorials},
  volume={26},
  number={3},
  pages={1520--1559},
  year={2024},
  publisher={IEEE}
}

@article{jeong2015random,
  title={Random access in millimeter-wave beamforming cellular networks: issues and approaches},
  author={Jeong, Cheol and others},
  journal={IEEE Communications Magazine},
  volume={53},
  number={1},
  pages={180--185},
  year={2015},
  publisher={IEEE}
}

@article{yang2023environment,
  title={Environment semantics aided wireless communications: A case study of mmWave beam prediction and blockage prediction},
  author={Yang, Yuwen and others},
  journal={IEEE journal on selected areas in communications},
  volume={41},
  number={7},
  pages={2025--2040},
  year={2023},
  publisher={IEEE}
}

@article{raha2025advancing,
  title={Advancing ultra-reliable 6 g: Transformer and semantic localization empowered robust beamforming in millimeter-wave communications},
  author={Raha, Avi Deb and others},
  journal={IEEE Transactions on Vehicular Technology},
  year={2025},
  publisher={IEEE}
}

@article{alkhateeb2014channel,
  title={Channel estimation and hybrid precoding for millimeter wave cellular systems},
  author={Alkhateeb, Ahmed and others},
  journal={IEEE journal of selected topics in signal processing},
  volume={8},
  number={5},
  pages={831--846},
  year={2014},
  publisher={IEEE}
}

@article{jayaprakasam2017robust,
  title={Robust beam-tracking for mmWave mobile communications},
  author={Jayaprakasam, Suhanya and others},
  journal={IEEE Communications Letters},
  volume={21},
  number={12},
  pages={2654--2657},
  year={2017},
  publisher={IEEE}
}

@article{heath2016overview,
  title={An overview of signal processing techniques for millimeter wave MIMO systems},
  author={Heath, Robert W and others},
  journal={IEEE journal of selected topics in signal processing},
  volume={10},
  number={3},
  pages={436--453},
  year={2016},
  publisher={IEEE}
}

@article{zhang2021learning,
  title={Learning reflection beamforming codebooks for arbitrary RIS and non-stationary channels},
  author={Zhang, Yu and others},
  journal={arXiv preprint arXiv:2109.14909},
  year={2021}
}

@inproceedings{sun2025towards,
  title={Towards Energy-Efficient Holographic MIMO Communications via Stacked Metasurface-Assisted Semantic Beamforming},
  author={Sun, Yifu and others},
  booktitle={ICC 2025-IEEE International Conference on Communications},
  pages={01--07},
  year={2025},
  organization={IEEE}
}

@article{sun2023define,
  title={How to define the propagation environment semantics and its application in scatterer-based beam prediction},
  author={Sun, Yutong and others},
  journal={IEEE Wireless Communications Letters},
  volume={12},
  number={4},
  pages={649--653},
  year={2023},
  publisher={IEEE}
}

@article{xie2024robust,
  title={Robust image semantic coding with learnable CSI fusion masking over MIMO fading channels},
  author={Xie, Bingyan and others},
  journal={IEEE Transactions on Wireless Communications},
  volume={23},
  number={10},
  pages={14155--14170},
  year={2024},
  publisher={IEEE}
}

@inproceedings{khan2025semqnet,
  title={SemQNet: Semantic-Aware Quantised Network for mmWave Beam Prediction},
  author={Khan, Ahsan Raza and others},
  booktitle={2025 IEEE Wireless Communications and Networking Conference (WCNC)},
  pages={1--6},
  year={2025},
  organization={IEEE}
}

@article{wen2023vision,
  title={Vision aided environment semantics extraction and its application in mmWave beam selection},
  author={Wen, Feiyang and others},
  journal={IEEE Communications Letters},
  volume={27},
  number={7},
  pages={1894--1898},
  year={2023},
  publisher={IEEE}
}

@article{habibi2019comprehensive,
  title={A comprehensive survey of RAN architectures toward 5G mobile communication system},
  author={Habibi, Mohammad Asif and others},
  journal={Ieee Access},
  volume={7},
  pages={70371--70421},
  year={2019},
  publisher={IEEE}
}

@article{zhang2025beamforming,
  title={Beamforming design for semantic-bit coexisting communication system},
  author={Zhang, Maojun and others},
  journal={IEEE Journal on Selected Areas in Communications},
  year={2025},
  publisher={IEEE}
}

@article{wu2024deep,
  title={Deep joint semantic coding and beamforming for near-space airship-borne massive MIMO network},
  author={Wu, Minghui and others},
  journal={IEEE Journal on Selected Areas in Communications},
  year={2024},
  publisher={IEEE}
}

@inproceedings{sheng2024semantic,
title={Semantic Communication for Cooperative Perception with HARQ},
author={Sheng, Yucheng and others},
booktitle={2024 IEEE 34th International Workshop on Machine Learning for Signal Processing (MLSP)},
pages={1--6},
year={2024},
organization={IEEE}
}

@article{zhou2022adaptive,
title={Adaptive bit rate control in semantic communication with incremental knowledge-based HARQ},
author={Zhou, Qingyang and others},
journal={IEEE Open Journal of the Communications Society},
volume={3},
pages={1076--1089},
year={2022},
publisher={IEEE}
}

@article{zheng2025semantic2,
title={Semantic Base Enabled Image Transmission With Fine-Grained HARQ},
author={Zheng, Yuan and others},
journal={IEEE Transactions on Wireless Communications},
year={2025},
publisher={IEEE}
}

@article{jiang2022deep,
title={Deep source-channel coding for sentence semantic transmission with HARQ},
author={Jiang, Peiwen and others},
journal={IEEE transactions on communications},
volume={70},
number={8},
pages={5225--5240},
year={2022},
publisher={IEEE}
}

@inproceedings{delfani2024semantics,
  title={Semantics-aware status updates with energy harvesting devices: Query version age of information},
  author={Delfani, Erfan and others},
  booktitle={2024 22nd International Symposium on Modeling and Optimization in Mobile, Ad Hoc, and Wireless Networks (WiOpt)},
  pages={177--184},
  year={2024},
  organization={IEEE}
}

@article{maatouk2020age,
  title={The age of incorrect information: A new performance metric for status updates},
  author={Maatouk, Ali and others},
  journal={IEEE/ACM Transactions on Networking},
  volume={28},
  number={5},
  pages={2215--2228},
  year={2020},
  publisher={IEEE}
}

@article{chiariotti2022query,
  title={Query age of information: Freshness in pull-based communication},
  author={Chiariotti, Federico and others},
  journal={IEEE Transactions on Communications},
  volume={70},
  number={3},
  pages={1606--1622},
  year={2022},
  publisher={IEEE}
}

@inproceedings{yates2021age,
  title={The age of gossip in networks},
  author={Yates, Roy D},
  booktitle={2021 IEEE International Symposium on Information Theory (ISIT)},
  pages={2984--2989},
  year={2021},
  organization={IEEE}
}

@inproceedings{kaul2012real,
  title={Real-time status: How often should one update?},
  author={Kaul, Sanjit and others},
  booktitle={2012 Proceedings IEEE INFOCOM},
  pages={2731--2735},
  year={2012},
  organization={IEEE}
}

@article{bourtsoulatze2019deep,
  title={Deep joint source-channel coding for wireless image transmission},
  author={Bourtsoulatze, Eirina and others},
  journal={IEEE Transactions on Cognitive Communications and Networking},
  volume={5},
  number={3},
  pages={567--579},
  year={2019},
  publisher={IEEE}
}

@article{li2024toward1,
  title={Toward goal-oriented semantic communications: New metrics, framework, and open challenges},
  author={Li, Aimin and others},
  journal={IEEE Wireless Communications},
  volume={31},
  number={5},
  pages={238--245},
  year={2024},
  publisher={IEEE}
}

@article{cao2025channel,
  title={Channel Code-Book (CCB): Semantic Image-Adaptive Transmission in Satellite--Ground Scenario},
  author={Cao, Hui and others},
  journal={Sensors},
  volume={25},
  number={1},
  pages={269},
  year={2025},
  publisher={MDPI}
}

@inproceedings{bian2025multi,
  title={Multi-Task Semantic Communication for Remote Sensing},
  author={Bian, Qing and others},
  booktitle={2025 IEEE 5th International Conference on Computer Communication and Artificial Intelligence (CCAI)},
  pages={70--75},
  year={2025},
  organization={IEEE}
}

@article{zhang2025flexible,
  title={Flexible Bit and Semantic on-demand Transmission Framework in Hyper-Reliable and Low Latency Communications Scenarios},
  author={Zhang, Jingxuan and others},
  journal={IEEE Transactions on Wireless Communications},
  year={2025},
  publisher={IEEE}
}

@article{du2024enabling,
  title={Enabling AI-generated content services in wireless edge networks},
  author={Du, Hongyang and others},
  journal={IEEE Wireless Communications},
  volume={31},
  number={3},
  pages={226--234},
  year={2024},
  publisher={IEEE}
}

@article{wang2023adaptive,
  title={Adaptive semantic-bit communication for extended reality interactions},
  author={Wang, Chaowei and others},
  journal={IEEE Journal of Selected Topics in Signal Processing},
  volume={17},
  number={5},
  pages={1080--1092},
  year={2023},
  publisher={IEEE}
}

@inproceedings{balakrishnan2025ontology,
  title={Ontology-Driven Semantic Interoperability Framework for Multi-Center Healthcare Big Data with Graph-Based Query Analysis},
  author={Balakrishnan, T Suresh and others},
  booktitle={2025 International Conference on Computing for Sustainability and Intelligent Future (COMP-SIF)},
  pages={1--6},
  year={2025},
  organization={IEEE}
}

@article{meng2024semantics,
  title={Semantics-empowered space-air-ground-sea integrated network: New paradigm, frameworks, and challenges},
  author={Meng, Siqi and others},
  journal={IEEE Communications Surveys \& Tutorials},
  volume={27},
  number={1},
  pages={140--183},
  year={2024},
  publisher={IEEE}
}

@inproceedings{hu2021manufacturing,
  title={Manufacturing Resource Semantic Modeling \& Description Towards Virtual Reorganization of Production Line Based on the IIoT},
  author={Hu, Hengwen and others},
  booktitle={2021 26th International Conference on Automation and Computing (ICAC)},
  pages={1--6},
  year={2021},
  organization={IEEE}
}

@inproceedings{zhu2023semantic,
  title={Semantic Reliability Maximization: A Cooperative Perspective in Integrated Sensing, Communication and Computation Networks},
  author={Zhu, Yao and others},
  booktitle={GLOBECOM 2023-2023 IEEE Global Communications Conference},
  pages={5073--5079},
  year={2023},
  organization={IEEE}
}

@Inbook{mao2021multiple,
author="Mao, Yijie
and Clerckx, Bruno",
editor="Lin, Xingqin
and Lee, Namyoon",
title="Multiple Access Techniques",
bookTitle="5G and Beyond: Fundamentals and Standards",
year="2021",
publisher="Springer International Publishing",
address="Cham",
pages="63--100",
isbn="978-3-030-58197-8",
doi="10.1007/978-3-030-58197-8_3",
url="https://doi.org/10.1007/978-3-030-58197-8_3"
}

@ARTICLE{zhang2023model,
  author={Ping, Zhang and others},
  journal={Frontiers of Information Technology \& Electronic Engineering}, 
  title={Model division multiple access for semantic communications}, 
  year={2023},
  volume={24},
  number={6},
  pages={801-812},
  doi={10.1109/JIOT.2024.3438082}}

@Inbook{frauendorf2023the,
author="Frauendorf, Jos{\'e} Luiz
and Almeida de Souza, {\'E}rika",
title="The Evolution of RAN (Radio Access Network), D-RAN, C-RAN, V-RAN, and O-RAN",
bookTitle="The Architectural and Technological Revolution of 5G",
year="2023",
publisher="Springer International Publishing",
address="Cham",
pages="139--154",
isbn="978-3-031-10650-7",
doi="10.1007/978-3-031-10650-7_10",
url="https://doi.org/10.1007/978-3-031-10650-7_10"
}

@INPROCEEDINGS{wu2025joint,
  author={Wu, Di and others},
  booktitle={2025 IEEE Wireless Communications and Networking Conference (WCNC)}, 
  title={Joint Deep Adversarial Semantic Decomposition Scheme for Model Division Multiple Access in IoT}, 
  year={2025},
  volume={},
  number={},
  pages={01-06},
  doi={10.1109/WCNC61545.2025.10978349}}

@ARTICLE{zhang2024deepma,
  author={Zhang, Wenyu and others},
  journal={IEEE Transactions on Cognitive Communications and Networking}, 
  title={DeepMA: End-to-End Deep Multiple Access for Wireless Image Transmission in Semantic Communication}, 
  year={2024},
  volume={10},
  number={2},
  pages={387-402},
  doi={10.1109/TCCN.2023.3326302}}

@INPROCEEDINGS{wu2023fusion,
  author={Wu, Tong and others},
  booktitle={GLOBECOM 2023 - 2023 IEEE Global Communications Conference}, 
  title={Fusion-Based Multi-User Semantic Communications for Wireless Image Transmission Over Degraded Broadcast Channels}, 
  year={2023},
  volume={},
  number={},
  pages={7623-7628},
  doi={10.1109/GLOBECOM54140.2023.10437864}}

@ARTICLE{liang2024orthogonal,
  author={Liang, Haotai and others},
  journal={IEEE Transactions on Wireless Communications}, 
  title={Orthogonal Model Division Multiple Access}, 
  year={2024},
  volume={23},
  number={9},
  pages={11693-11707},
  doi={10.1109/TWC.2024.3384421}}

@article{DeepLearningEmpoweredSharedKB,
  title={Deep learning-empowered semantic communication systems with a shared knowledge base},
  author={Yi, Peng and others},
  journal={IEEE Transactions on Wireless Communications},
  volume={23},
  number={6},
  pages={6174--6187},
  year={2023},
  publisher={IEEE}
}

@inproceedings{EndToEndGenerativeSKB,
  title={End-to-end generative semantic communication powered by shared semantic knowledge base},
  author={Li, Shuling and others},
  booktitle={2024 IEEE International Conference on Communications Workshops (ICC Workshops)},
  pages={1067--1072},
  year={2024},
  organization={IEEE}
}

@article{KnowledgeEnhancedReceiver,
  title={Knowledge enhanced semantic communication receiver},
  author={Wang, Bingyan and others},
  journal={IEEE Communications Letters},
  volume={27},
  number={7},
  pages={1794--1798},
  year={2023},
  publisher={IEEE}
}

@article{CognitiveSemComKG,
  title={Cognitive semantic communication systems driven by knowledge graph: Principle, implementation, and performance evaluation},
  author={Zhou, Fuhui and others},
  journal={IEEE Transactions on Communications},
  volume={72},
  number={1},
  pages={193--208},
  year={2023},
  publisher={IEEE}
}

@article{ExplicitSemanticBase,
  title={Explicit Semantic-Base-Empowered Communications for 6G Mobile Networks},
  author={Wang, Fengyu and others},
  journal={Engineering},
  year={2025},
  publisher={Elsevier}
}

@inproceedings{UnifiedHierarchicalSKB,
  title={A unified hierarchical semantic knowledge base for multi-task semantic communication},
  author={Wang, Lingyi and others},
  booktitle={ICC 2024-IEEE International Conference on Communications},
  pages={2937--2943},
  year={2024},
  organization={IEEE}
}

@article{KnowledgeBaseEnabledGenerative,
  title={Knowledge base enabled semantic communication: A generative perspective},
  author={Ren, Jinke and others},
  journal={IEEE Wireless Communications},
  volume={31},
  number={4},
  pages={14--22},
  year={2024},
  publisher={IEEE}
}

@article{lu2024semantic,
  title={Semantic-aware vision-assisted integrated sensing and communication: Architecture and resource allocation},
  author={Lu, Yang and others},
  journal={IEEE Wireless Communications},
  volume={31},
  number={3},
  pages={302--308},
  year={2024},
  publisher={IEEE}
}

@article{xu2025generative,
  title={Generative semantic communications with foundation models: Perception-error analysis and semantic-aware power allocation},
  author={Xu, Chunmei and others},
  journal={IEEE Journal on Selected Areas in Communications},
  year={2025},
  publisher={IEEE}
}

@article{ding2023joint,
  title={Joint urllc traffic scheduling and resource allocation for semantic communication systems},
  author={Ding, Guangyao and others},
  journal={IEEE Transactions on Wireless Communications},
  volume={23},
  number={7},
  pages={7278--7290},
  year={2023},
  publisher={IEEE}
}

@article{zheng2024semantic,
  title={Semantic communication in satellite-borne edge cloud network for computation offloading},
  author={Zheng, Guhan and others},
  journal={IEEE Journal on Selected Areas in Communications},
  volume={42},
  number={5},
  pages={1145--1158},
  year={2024},
  publisher={IEEE}
}

@article{wang2024adaptive,
  title={Adaptive resource allocation for semantic communication networks},
  author={Wang, Lingyi and others},
  journal={IEEE Transactions on Communications},
  volume={72},
  number={11},
  pages={6900--6916},
  year={2024},
  publisher={IEEE}
}

@article{xia2023joint,
  title={Joint user association and bandwidth allocation in semantic communication networks},
  author={Xia, Le and others},
  journal={IEEE Transactions on Vehicular Technology},
  volume={73},
  number={2},
  pages={2699--2711},
  year={2023},
  publisher={IEEE}
}

@article{cang2023online,
  title={Online resource allocation for semantic-aware edge computing systems},
  author={Cang, Yihan and others},
  journal={IEEE Internet of Things Journal},
  volume={11},
  number={17},
  pages={28094--28110},
  year={2023},
  publisher={IEEE}
}

@inproceedings{zhao2023joint,
  title={Joint computing resource and bandwidth allocation for semantic communication networks},
  author={Zhao, Fangzhou and others},
  booktitle={2023 IEEE 98th Vehicular Technology Conference (VTC2023-Fall)},
  pages={1--5},
  year={2023},
  organization={IEEE}
}

@article{liu2023adaptable,
  title={Adaptable semantic compression and resource allocation for task-oriented communications},
  author={Liu, Chuanhong and others},
  journal={IEEE Transactions on Cognitive Communications and Networking},
  volume={10},
  number={3},
  pages={769--782},
  year={2023},
  publisher={IEEE}
}

@inproceedings{hu2023semantic,
  title={Semantic-oriented resource allocation for multi-modal UAV semantic communication networks},
  author={Hu, Han and others},
  booktitle={GLOBECOM 2023-2023 IEEE Global Communications Conference},
  pages={7213--7218},
  year={2023},
  organization={IEEE}
}

@inproceedings{minani2024qosem,
  title={QoSem-based Resource Allocation for Semantic Communication in NTN Downlinks},
  author={Minani, Frodouard and others},
  booktitle={GLOBECOM 2024-2024 IEEE Global Communications Conference},
  pages={2587--2592},
  year={2024},
  organization={IEEE}
}

@article{hu2025resource,
  title={Resource allocation for multi-modal semantic communication in UAV collaborative networks},
  author={Hu, Han and others},
  journal={IEEE Transactions on Communications},
  year={2025},
  publisher={IEEE}
}

@article{adhikary2024holographic,
  title={Holographic MIMO with integrated sensing and communication for energy-efficient cell-free 6G networks},
  author={Adhikary, Apurba and others},
  journal={IEEE Internet of Things Journal},
  volume={11},
  number={19},
  pages={30617--30635},
  year={2024},
  publisher={IEEE}
}

@article{LessDataMoreKnowledge,
  title={Less data, more knowledge: Building next-generation semantic communication networks},
  author={Chaccour, Christina and others},
  journal={IEEE Communications Surveys \& Tutorials},
  volume={27},
  number={1},
  pages={37--76},
  year={2024},
  publisher={IEEE}
}

@article{yang2024secure,
  title={Secure semantic communications: Fundamentals and challenges},
  author={Yang, Zhaohui and others},
  journal={IEEE network},
  volume={38},
  number={6},
  pages={513--520},
  year={2024},
  publisher={IEEE}
}

@inproceedings{tang2023gan,
  title={GAN-inspired intelligent jamming and anti-jamming strategy for semantic communication systems},
  author={Tang, Rui and others},
  booktitle={2023 IEEE International Conference on Communications Workshops (ICC Workshops)},
  pages={1623--1628},
  year={2023},
  organization={IEEE}
}

@article{shen2023secure,
  title={Secure semantic communications: Challenges, approaches, and opportunities},
  author={Shen, Meng and others},
  journal={IEEE Network},
  volume={38},
  number={4},
  pages={197--206},
  year={2023},
  publisher={IEEE}
}

@inproceedings{do2025security,
  title={Security and Privacy Challenges in Semantic Communication Networks},
  author={Do, Quang Tuan and others},
  booktitle={2025 International Conference on Artificial Intelligence in Information and Communication (ICAIIC)},
  pages={0032--0035},
  year={2025},
  organization={IEEE}
}

@article{peng2024adversarial,
  title={Adversarial reinforcement learning based data poisoning attacks defense for task-oriented multi-user semantic communication},
  author={Peng, Jincheng and others},
  journal={IEEE Transactions on Mobile Computing},
  year={2024},
  publisher={IEEE}
}

@inproceedings{sagduyu2023vulnerabilities,
  title={Vulnerabilities of deep learning-driven semantic communications to backdoor (trojan) attacks},
  author={Sagduyu, Yalin E and others},
  booktitle={2023 57th Annual Conference on Information Sciences and Systems (CISS)},
  pages={1--6},
  year={2023},
  organization={IEEE}
}

@article{wang2025diffusion,
  title={Diffusion-based Task-oriented Semantic Communications with Model Inversion Attack},
  author={Wang, Xuesong and others},
  journal={arXiv preprint arXiv:2506.19886},
  year={2025}
}

@article{zhang2023swinjscc,
  title={SwinJSCC: Taming Swin transformer for deep joint source-channel coding},
  author={Zhang, Ke and others},
  journal={IEEE Journal on Selected Areas in Communications},
  volume={41},
  number={8},
  pages={2619--2634},
  year={2023},
  publisher={IEEE}
}

@article{dai2022nonlinear,
  title={Nonlinear transform source-channel coding for semantic communications},
  author={Dai, Jincheng and others},
  journal={IEEE Journal on Selected Areas in Communications},
  volume={40},
  number={8},
  pages={2300--2316},
  year={2022},
  publisher={IEEE}
}

@article{bo2024joint,
  title={Joint coding-modulation for digital semantic communications via variational autoencoder},
  author={Bo, Yufei and others},
  journal={IEEE Transactions on Communications},
  volume={72},
  number={9},
  pages={5626--5640},
  year={2024},
  publisher={IEEE}
}

@article{lokumarambage2023wireless,
  title={Wireless end-to-end image transmission system using semantic communications},
  author={Lokumarambage, Maheshi U and others},
  journal={IEEE Access},
  volume={11},
  pages={37149--37163},
  year={2023},
  publisher={IEEE}
}

@article{salehi2025llm,
  title={Llm-enabled data transmission in end-to-end semantic communication},
  author={Salehi, Shavbo and others},
  journal={arXiv preprint arXiv:2504.07431},
  year={2025}
}

@article{meng2025secure,
  title={Secure semantic communication with homomorphic encryption},
  author={Meng, Rui and others},
  journal={arXiv preprint arXiv:2501.10182},
  year={2025}
}

@article{liu2023semprotector,
  title={SemProtector: A unified framework for semantic protection in deep learning-based semantic communication systems},
  author={Liu, Xinghan and others},
  journal={IEEE Communications Magazine},
  volume={61},
  number={11},
  pages={56--62},
  year={2023},
  publisher={IEEE}
}

@article{fan2025generative,
  title={Generative diffusion models for wireless networks: Fundamental, architecture, and state-of-the-art},
  author={Fan, Dayu and others},
  journal={arXiv preprint arXiv:2507.16733},
  year={2025}
}

@article{kang2023adversarial,
  title={Adversarial attacks and defenses for semantic communication in vehicular metaverses},
  author={Kang, Jiawen and others},
  journal={IEEE Wireless Communications},
  volume={30},
  number={4},
  pages={48--55},
  year={2023},
  publisher={IEEE}
}

@article{cao2025importance,
  title={Importance-aware robust semantic transmission for leo satellite-ground communication},
  author={Cao, Hui and others},
  journal={arXiv preprint arXiv:2508.11457},
  year={2025}
}

@inproceedings{chen2024enhancing,
  title={Enhancing image privacy in semantic communication over wiretap channels leveraging differential privacy},
  author={Chen, Weixuan and others},
  booktitle={2024 IEEE 34th International Workshop on Machine Learning for Signal Processing (MLSP)},
  pages={1--6},
  year={2024},
  organization={IEEE}
}

@article{feng2025towards,
  title={Towards 6G Native-AI Edge Networks: A Semantic-Aware and Agentic Intelligence Paradigm},
  author={Feng, Chenyuan and others},
  journal={arXiv preprint arXiv:2512.04405},
  year={2025}
}

@article{fan2025kgrag,
  title={KGRAG-SC: Knowledge Graph RAG-Assisted Semantic Communication},
  author={Fan, Dayu and others},
  journal={arXiv preprint arXiv:2509.04801},
  year={2025}
}

@article{nguyen2024distortion,
  title={Distortion resilience for goal-oriented semantic communication},
  author={Nguyen, Minh-Duong and others},
  journal={IEEE Transactions on Mobile Computing},
  volume={24},
  number={5},
  pages={3489--3501},
  year={2024},
  publisher={IEEE}
}

@inproceedings{liu2024adaptive,
  title={Adaptive privacy budget-based differential privacy co-training for wireless semantic communication},
  author={Liu, Honghao and others},
  booktitle={2024 IEEE Wireless Communications and Networking Conference (WCNC)},
  pages={1--6},
  year={2024},
  organization={IEEE}
}

@inproceedings{yuan2025homomorphic,
  title={Homomorphic Encryption-Based Joint Source-Channel Coding for Semantic Communications},
  author={Yuan, Yifan and others},
  booktitle={2025 IEEE 26th International Workshop on Signal Processing and Artificial Intelligence for Wireless Communications (SPAWC)},
  pages={1--5},
  year={2025},
  organization={IEEE}
}

@article{gao2025semstediff,
  title={SemSteDiff: Generative Diffusion Model-based Coverless Semantic Steganography Communication},
  author={Gao, Song and others},
  journal={arXiv preprint arXiv:2509.04803},
  year={2025}
}

@inproceedings{jin2022ris,
  title={RIS-assisted physical layer key generation and transmit power minimization},
  author={Jin, Liang and others},
  booktitle={2022 IEEE Wireless Communications and Networking Conference (WCNC)},
  pages={2065--2070},
  year={2022},
  organization={IEEE}
}

@article{gao2023esanet,
  title={EsaNet: Environment semantics enabled physical layer authentication},
  author={Gao, Ning and others},
  journal={IEEE wireless communications letters},
  volume={13},
  number={1},
  pages={178--182},
  year={2023},
  publisher={IEEE}
}

@article{wang2025image,
  title={Image Steganography For Securing Intellicise Wireless Networks:" Invisible Encryption" Against Eavesdroppers},
  author={Wang, Bizhu and others},
  journal={arXiv preprint arXiv:2505.04467},
  year={2025}
}

@article{meng2022physical,
  title={Physical-layer authentication based on hierarchical variational autoencoder for Industrial Internet of Things},
  author={Meng, Rui and others},
  volume={10},
  number={3},
  pages={2528--2544},
  year={2023},
  publisher={IEEE}
}

@article{meng2022multiuser,
  title={Multiuser physical-layer authentication based on latent perturbed neural networks for industrial internet of things},
  author={Meng, Rui and others},
  journal={IEEE Internet of Things journal},
  volume={10},
  number={1},
  pages={637--652},
  year={2022},
  publisher={IEEE}
}

@article{meng2025generative,
  title={Generative AI for Physical-Layer Authentication},
  author={Meng, Rui and others},
  journal={arXiv preprint arXiv:2504.18175},
  year={2025}
}

@book{niu2025mathematical,
  title={The Mathematical Theory of Semantic Communication},
  author={Niu, Kai and Zhang, Ping},
  year={2025},
  publisher={Springer Nature}
}

@inproceedings{tan2024security,
  title={Security Improvement for Deep Learning-Based Semantic Communication Systems},
  author={Tan, Hung Ngo Luu and others},
  booktitle={2024 International Conference on Advanced Technologies for Communications (ATC)},
  pages={717--721},
  year={2024},
  organization={IEEE}
}

@article{zhou2025rome,
  title={ROME: Robust Model Ensembling for Semantic Communication Against Semantic Jamming Attacks},
  author={Zhou, Kequan and others},
  journal={arXiv preprint arXiv:2501.01172},
  year={2025}
}

@article{lin2025blockchain,
  title={Blockchain-based knowledge-aware semantic communications for remote driving image transmission.},
  author={Lin, Yangfei and others},
  journal={Digit. Commun. Networks},
  volume={11},
  number={1},
  pages={317--325},
  year={2025},
  publisher={Elsevier}
}

@inproceedings{zhou2024backdoor,
  title={Backdoor attacks and defenses on semantic-symbol reconstruction in semantic communications},
  author={Zhou, Yuan and others},
  booktitle={ICC 2024-IEEE International Conference on Communications},
  pages={734--739},
  year={2024},
  organization={IEEE}
}

@inproceedings{wang2024polling,
  title={Polling Resource Scheduling for Internet of Vehicles Based on U-Net Semantic Segmentation},
  author={Wang, Junjie and others},
  booktitle={2024 10th International Conference on Computer and Communications (ICCC)},
  pages={1619--1624},
  year={2024},
  organization={IEEE}
}

@article{zhou2024semantic,
  title={Semantic information extraction and multi-agent communication optimization based on generative pre-trained transformer},
  author={Zhou, Li and others},
  journal={IEEE Transactions on Cognitive Communications and Networking},
  year={2024},
  publisher={IEEE}
}

@article{tang2023cooperative,
  title={Cooperative semantic communication with on-demand semantic forwarding},
  author={Tang, Bing and others},
  journal={IEEE Open Journal of the Communications Society},
  volume={5},
  pages={349--363},
  year={2023},
  publisher={IEEE}
}

@inproceedings{hu2022robust,
  title={Robust semantic communications against semantic noise},
  author={Hu, Qiyu and others},
  booktitle={2022 IEEE 96th Vehicular Technology Conference (VTC2022-Fall)},
  pages={1--6},
  year={2022},
  organization={IEEE}
}

@article{yang2023gradient,
  title={Gradient leakage attacks in federated learning: Research frontiers, taxonomy, and future directions},
  author={Yang, Haomiao and others},
  journal={IEEE Network},
  volume={38},
  number={2},
  pages={247--254},
  year={2023},
  publisher={IEEE}
}

@article{zhou2024secure,
  title={Secure multi-party computation for machine learning: A survey},
  author={Zhou, Ian and others},
  journal={IEEE Access},
  volume={12},
  pages={53881--53899},
  year={2024},
  publisher={IEEE}
}

@article{jauernig2020trusted,
  title={Trusted execution environments: properties, applications, and challenges},
  author={Jauernig, Patrick and others},
  journal={IEEE Security \& Privacy},
  volume={18},
  number={2},
  pages={56--60},
  year={2020},
  publisher={IEEE}
}

@inproceedings{zhou2022survey,
  title={A survey of security aggregation},
  author={Zhou, Sufang and others},
  booktitle={2022 24th International Conference on Advanced Communication Technology (ICACT)},
  pages={334--340},
  year={2022},
  organization={IEEE}
}

@inproceedings{liu2025learning,
  title={Learning-based power control for secure covert semantic communication},
  author={Liu, Yansheng and others},
  booktitle={2025 International Wireless Communications and Mobile Computing (IWCMC)},
  pages={257--262},
  year={2025},
  organization={IEEE}
}

@article{meng2025survey2,
  title={A survey of machine learning-based physical-layer authentication in wireless communications},
  author={Meng, Rui and others},
  journal={Journal of network and computer applications},
  volume={235},
  pages={104085},
  year={2025},
  publisher={Elsevier}
}

@article{rodriguez2015physical,
  title={Physical layer security in wireless cooperative relay networks: State of the art and beyond},
  author={Rodriguez, Leonardo Jimenez and others},
  journal={IEEE Communications Magazine},
  volume={53},
  number={12},
  pages={32--39},
  year={2015},
  publisher={IEEE}
}

@article{gou2021knowledge,
  title={Knowledge distillation: A survey},
  author={Gou, Jianping and others},
  journal={International journal of computer vision},
  volume={129},
  number={6},
  pages={1789--1819},
  year={2021},
  publisher={Springer}
}

@article{zhang2025optimization,
  title={Optimization of Private Semantic Communication Performance: An Uncooperative Covert Communication Method},
  author={Zhang, Wenjing and others},
  journal={IEEE transactions on wireless communications},
  year={2025},
  publisher={IEEE}
}

@article{rizvi2025controlled,
  title={Controlled Quantum Semantic Communication for Industrial CPS Networks},
  author={Rizvi, Syed Muhammad Abuzar and others},
  journal={IEEE Transactions on Network Science and Engineering},
  year={2025},
  publisher={IEEE}
}

@inproceedings{dai2024secure,
  title={Secure resource allocation for integrated sensing and semantic communication system},
  author={Dai, Jianxin and others},
  booktitle={2024 IEEE International Conference on Communications Workshops (ICC Workshops)},
  pages={1225--1230},
  year={2024},
  organization={IEEE}
}

@article{zhousecurity,
  title={Security Optimization of Semantic Communication System Based on Reconfigurable Intelligent Surface},
  author={Zhou, Zhiling}
}

@article{lan2021survey,
  title={A survey on complex knowledge base question answering: Methods, challenges and solutions},
  author={Lan, Yunshi and others},
  journal={arXiv preprint arXiv:2105.11644},
  year={2021}
}

@inproceedings{chu2016data,
  title={Data cleaning: Overview and emerging challenges},
  author={Chu, Xu and others},
  booktitle={Proceedings of the 2016 international conference on management of data},
  pages={2201--2206},
  year={2016}
}

@inproceedings{zhao2022semkey,
  title={SemKey: Boosting secret key generation for RIS-assisted semantic communication systems},
  author={Zhao, Ran and others},
  booktitle={2022 IEEE 96th Vehicular Technology Conference (VTC2022-Fall)},
  pages={1--5},
  year={2022},
  organization={IEEE}
}

@article{meng2023multidimensional,
  title={Multidimensional fingerprints-based multiattacker detection for 6G systems},
  author={Meng, Rui and others},
  journal={IEEE Internet of Things Journal},
  volume={11},
  number={2},
  pages={2665--2683},
  year={2023},
  publisher={IEEE}
}

@article{zhang2025balancing,
  title={Balancing Security and Efficiency in GAI-Driven Semantic Communication: Challenges, Solutions, and Future Paths},
  author={Zhang, Qianyun and others},
  journal={IEEE Network},
  year={2025},
  publisher={IEEE}
}

@article{he2023structured,
  title={Structured pruning for deep convolutional neural networks: A survey},
  author={He, Yang and others},
  journal={IEEE transactions on pattern analysis and machine intelligence},
  volume={46},
  number={5},
  pages={2900--2919},
  year={2023},
  publisher={IEEE}
}

@article{swaminathan2020sparse,
  title={Sparse low rank factorization for deep neural network compression},
  author={Swaminathan, Sridhar and others},
  journal={Neurocomputing},
  volume={398},
  pages={185--196},
  year={2020},
  publisher={Elsevier}
}

@article{park2024vision,
  title={Vision transformer-based semantic communications with importance-aware quantization},
  author={Park, Joohyuk and others},
  journal={arXiv preprint arXiv:2412.06038},
  year={2024}
}

@misc{rajan2024data,
  title={Data cleaning for machine learning systems-A survey},
  author={Rajan, Amit},
  year={2024}
}
\bibliographystyle{IEEEtran}

\vfill

\end{document}